\documentclass{aa}  

\usepackage{graphicx}
\usepackage{xcolor}

\newcommand{\htwo}{H$_2$}

\newcommand{\oi}{[\ion{O}{i}]}

\newcommand{\kms}{km\,s$^{-1}$}
\newcommand{\um}{$\mu$m}

\newcommand{\lsun}{L$_{\odot}$}

\newcommand{\cmt}{cm$^{-3}$}

\usepackage{txfonts}
\usepackage{amstext}

\begin{document}

   \title{GIARPS High-resolution Observations of T Tauri stars (GHOsT)}

   \subtitle{II. Connecting atomic and molecular winds in protoplanetary disks}

   \author{Gangi, M.             \inst{1}
          \and Nisini, B.        \inst{1}
          \and Antoniucci, S.    \inst{1}
          \and Giannini, T.      \inst{1}
          \and Biazzo, K.        \inst{1}
          \and Alcal\'a, J. M.   \inst{2}
          \and Frasca, A.        \inst{3}
          \and Munari, U.        \inst{4}
          \and Arkharov, A. A.   \inst{5}
          \and Harutyunyan, A.   \inst{6}
          \and Manara, C.F.      \inst{7}
          \and Rigliaco, E.      \inst{8}
          \and Vitali, F.        \inst{1}
        }

   \institute{INAF - Osservatorio Astronomico di Roma - Via Frascati 33, 00078 Monte Porzio Catone, Italy \\
              \email{manuele.gangi@inaf.it}
        \and
             INAF - Osservatorio Astronomico di Capodimonte - Salita Moiariello 16, 80131 Napoli, Italy
        \and
             INAF - Osservatorio Astrofisico di Catania - Via S. Sofia 78, 95123 Catania, Italy
        \and
            INAF–Osservatorio Astronomico di Padova, via dell’Osservatorio 8, 36012 Asiago (VI), Italy
        \and
            Central Astronomical Observatory of Pulkovo, Pulkovskoe shosse 65, 196140 St. Petersburg, Russia
        \and
             Fundaci\'on Galileo Galilei - INAF - Telescopio Nazionale Galileo, 38700 Bre$\rm \tilde n$a Baja, Santa Cruz de Tenerife, Spain
        \and
            European Southern Observatory, Karl-Schwarzschild-Strasse 2, 85748 Garching bei M\"unchen, Germany
        \and
            INAF–Osservatorio Astronomico di Padova, vicolo dell’ Osservatorio 5, 35122, Padova, Italy
             }

   \date{Received Month XX, XXXX; accepted Month XX, XXXX}

 
  \abstract
   {}
   {In the framework of the GIARPS High-resolution Observations of T Tauri stars (GHOsT) project, we aim to characterize the atomic and molecular winds in a sample of classical T Tauri stars (CTTs) of the Taurus-Auriga region.}
   {We analyzed the flux calibrated \oi\ 630 nm and \htwo\ 2.12 $\rm \mu m$ lines in a sample of 36 CTTs observed at the Telescopio Nazionale Galileo with the HARPS and GIANO spectrographs. We decomposed the line profiles into different kinematic Gaussian components and focused on the most frequently detected component, the narrow low-velocity (v$\rm_p < 20$ $\rm km$ $\rm s^{-1}$) component (NLVC).}
   {We found that the \htwo\ line is detected in 17 sources ($\sim 50 \%$ detection rate), and \oi\ is detected in all sources but one. The NLV components of the \htwo\ and \oi\ emission are kinematically linked, with a strong correlation between the peak velocities and the full widths at half maximum of the two lines. Assuming Keplerian broadening, we found that the \oi\ NVLC originates from a disk region between 0.05 and 20 au and that of \htwo\ in a region from 2 and 20 au. We did not find any clear correlation between v$\rm_p$ of the \htwo\ and \oi\ NVLC and the outer disk inclination. This result is in line with previous studies.}
   {Our results suggest that molecular and neutral atomic emission in disk winds originate from regions that might overlap, and that the survival of molecular winds in disks strongly depends on the gas exposure to the radiation from the central star. Our results demonstrate the potential of wide-band high-resolution spectroscopy in linking tracers of different manifestations of the same phenomenon.}


   \maketitle
%

\section{Introduction}

Disk winds are a key ingredient in the evolution of protoplanetary systems. Together with accretion, they are responsible for gas dissipation in the inner (i.e., < 10 au) disk region \citep{Ercolano2017}. In addition, magnetic disk winds play a fundamental role in the extraction of angular momentum, thus allowing matter to be transferred from the outer to the inner disk region and then onto the central star. Winds also alter the disk surface density and eventually affect the process that leads to the formation and migration of protoplanets \citep{Suzuki2016}.  
Two main classes of winds have been proposed to work in low-mass pre-main-sequence stars: i) magnetohydrodynamically (MHD) accelerated winds, including the high-velocity collimated jets originating in the inner disk region, which drives accretion by removing angular momentum \citep[e.g.,][]{Ferreira2006,Turner2014,Bai2016}; and ii) photoevaporative pressure-driven winds, caused by the effect of UV and X-ray photons incident on the disk surface \citep{Alexander2006, Ercolano2008}. Photoevaporative winds do not play a role in the extraction of angular momentum, but are believed to have a strong effect at later stages of disk evolution by accelerating the disk dissipation \citep{Alexander2014}.

In classical T Tauri (CTT) stars, jets and winds are studied through atomic or weakly ionized forbidden lines in the optical/IR spectral range. Among the main tracers of atomic winds, the [\ion{O}{I}] $\rm 630$ $\rm nm$ line is the brightest and therefore the most frequently studied line to infer disk-wind properties since the pioneering spectroscopic works of \citet{Hartigan1995} and \citet{Hirth1997}.
Optical observations at moderate and high spectral resolution have shown that the \oi\ line has a composite profile that traces distinct physical components. Emission with a large (positive or negative) shift with respect to the systemic velocity (the high-velocity component, HVC) is associated with collimated extended jets. In addition, \oi\ exhibits components at lower velocity (v$\rm_p \la 30$ $\rm km$ $\rm s^{-1}$) that today are attributed to extended ($\rm 0.5-10$ $\rm au$) disk winds. Excitation analysis of this low-velocity component (LVC) has revealed that it is due to a warm and dense gas ($\rm T< 10,000$ $\rm K$, $\rm n_{tot}$ $\simeq$ $10^6$ - $10^7$ cm$^{-3}$) at an ionization rate $\rm x_e$ lower than the high-velocity emission \citep[$\rm x_e < 0.1$,][]{Giannini2019, Natta2014}. The luminosity of the LVC and HVC correlates with the accretion luminosity and mass accretion rates with a similar slope \citep{Nisini2018}. This suggests that the luminosity of the two components is due to the heating from the accretion process. On the other hand, it implies that the two components of the lines are not necessarily emitted in the same wind \citep{Weber2020}. In some cases, when the [\ion{O}{I}] emission is observed
at high spectral resolution, it displays more than one component at low velocity. If represented with Gaussian functions, the most frequently detected component is  narrow (full width at half maximum, $\rm FWHM < 40$ $\rm km$ $\rm s^{-1}$), centrally peaked, or only slightly blueshifted with respect to the source velocity. In addition to this, a broader component is often observed (BLVC) (\citealt{Rigliaco2013, Simon2016, Mcginnis2018, Banzatti2019}).  Based on the relative detection frequency of the two components, it has been suggested that the LV narrow component persists for longer times in the disk evolution because it has also been detected, at variance with the others, in more evolved sources with evidence of optically thin and dust-depleted inner disks \citep{Banzatti2019}. 
Correlations found between the LVC and HVC also suggest that MHD winds are at the origin of both emissions.
However, photoevaporative winds (e.g., \citealt{Ercolano2016}) might also play a role in driving the narrow LVC, at least for the most evolved sources.

Together with forbidden atomic emission, spectra of T Tauri stars in the IR often display emission from ro-vibrational transitions of molecular hydrogen, the 1-0 S(1) line at $\rm 2.12$ $\rm \mu m$ in particular. The $\rm H_2$ $\rm 2.12$ $ \rm \mu m$ line, when observed with high spectral resolution, shows a width of about $10-20$ $\rm km\ s^{-1}$ and the velocity peak is slightly blueshifted ($< 10$ $\rm km$ $\rm s^{-1}$) or compatible with systemic velocity. Spectral imaging observations with high spatial resolution have shown a variety of morphologies for the $\rm H_2$ emission region, from extended poorly collimated winds to more compact regions that are compatible with a gaseous disk origin \citep{Bary2003,Beck2008,Garcia2013,Agra-Amboage2014,Beck2019}. Excitation mechanisms for the $\rm H_2$ emission are not well constrained and might include low-velocity nondissociative shocks due to the interaction of the atomic wind with the surrounding disk or envelope material, photon irradiation of the disk atmosphere, or a molecular disk wind heated by ambipolar diffusion \citep[e.g.,][]{Panoglou2012, Agra-Amboage2014}. Multiple excitation mechanisms are likely at play in the complex environment of the star-disk region.

When the kinematic properties of the $\rm H_2$ $\rm 2.12$ $\rm \mu m$ and [\ion{O}{I}] $630$ $\rm nm$ lines are compared, it can be noted that the $\rm H_2$ line profile is broadly similar to the narrow component of the LV [\ion{O}{I}] emission. They share similar widths ($10-30$ $\rm km$ $\rm s^{-1}$) and the same slightly blueshifted line peaks. However, these two emissions have rarely been analyzed together, and studies connecting the atomic and molecular gas emitted in protoplanetary disk winds are lacking so far. \citet{Agra-Amboage2014} compared the $\rm H_2$ kinematics with that of the [\ion{O}{I}] in DG Tau, suggesting that $\rm H_2$ traces a wider and less collimated wind component than [\ion{O}{I}], in agreement with what is expected in MHD disk-wind models. However, large uncertainties remain when optical and near-IR data taken in different epochs and with different instrument configurations are compared because effects such as the line variability can affect the interpretation.

An important step forward for this type of studies can be made by using spectroscopic instrumentation that simultaneously covers the optical and IR spectral range at high resolution. In this respect, the GIARPS observing mode \citep{Claudi2017} of the Telescopio Nazionale Galileo (TNG) (La Palma, Spain), which takes advantage of the simultaneous use of the high-resolution spectrographs HARPS-N (optical) and GIANO-B (near-IR), is today the most well-suited facility to obtain simultaneous observations at high spectral dispersion of the $\rm H_2$ $\rm 2.12$ $\rm \mu m$ and the [\ion{O}{I}] $\rm 630$ $\rm nm$ lines in relatively bright sources. 

In this paper we report on the study of the relation between the atomic and molecular emission in a sample of 36 classical T Tauri stars (CTTs) using GIARPS spectra acquired within the framework of the GIARPS High-resolution Observations of T Tauri stars (GHOsT) project, which aims to characterize the properties of CTTs of the Taurus-Auriga region by combining high-resolution optical and IR spectroscopy. A first paper of the GHOsT project \citep{Giannini2019} has addressed the kinematic and physical properties of the jets and winds in a limited sample of six sources through a diagnostic analysis that combines ratios of optical and IR lines. In this work, we expand our observed sample and focus on a statistical analysis of the kinematic properties of the \oi\ $\rm 630$ $\rm nm$ and $\rm H_2$ $\rm 2.12$ $\rm \mu m$ lines and their mutual relation.

The paper is organized as follows. In Sections 2 and 3 we describe the observations and reduction of the GIARPS data together with ancillary data taken at different facilities that are used to flux-calibrate the GIARPS spectra. In Section 4 we describe the decomposition performed on the \oi\ and \htwo\ line profiles in order to identify the different kinematic components. In this section we also discuss the variability of the line profiles and any correlation between the kinematic properties and the disk inclination. In Section 5 we compare our results with previous studies on the \oi\ and \htwo\ emission in T Tauri stars, and discuss the implications of our findings in the framework of existing models.

\section{Observations}
The GHOsT targets were selected based on the Taurus-Auriga young population census by \citet{Esplin2014}, considering sources with $J <$ 11 mag and $R <$ 13.5 mag. In this paper we discuss 36 of these sources, which represent 34\% of the entire sample. The observed sources span masses between 0.1 and 2 $\rm M_{\sun}$, which is representative of the distribution in the complete sample. The logbook of observations is given in Table \ref{tab:Logbook}, while Table \ref{tab:SourceProperties} provides the most relevant parameters of the targets: distance, spectral type, visual extinction, luminosity, mass, and disk inclination. The latter were retrieved from the most recent literature. Details about the GIARPS observations and about the ancillary low-resolution spectroscopy and photometric data that were used to flux-calibrate the spectra are given in the following sections.

\subsection{GIARPS observations}
We extended the GIARPS observations made in 2017 that we reported in \cite{Giannini2019} with new observations taken in 2018 (December 8-9), 2019 (November 1-2), and 2020 (January 25-26). In addition, we also included spectra acquired in the IR range alone, taken with the GIANO-A instrument in 2015 and 2016. A total of 34 objects was then observed both in the IR and optical range, while 2 objects (DD Tau and HL Tau) were observed in the IR alone.

As already mentioned, the GIARPS mode allows the simultaneous use of the HARPS-N and GIANO-B spectrographs.
HARPS-N is an optical ($\rm 390-690$ $\rm nm$) echelle spectrograph with a resolving power of $\rm R=115000$ \citep{Cosentino2012}. It is fed by two fibres from the Nasmyth B focus of the TNG telescope with a field of view ($\rm FoV) \text{ of }1''$. For each observation, a fiber was placed on the target and the other was placed on the sky. The total exposure times are reported in Table \ref{tab:Logbook}.

GIANO-B is a near-infrared ($\rm 950-2400$ $\rm nm$) cross-dispersed echelle spectrograph with a resolving power of $\rm R=50000$ \citep{Oliva2012, Origlia2014}. The instrument slit has on-sky dimensions of $\rm 6''\times 0.5''$ and is directly fed by the telescope light. The spectra were acquired with the nodding A-B strategy, which consists of alternately observing the object in predefined "A" and "B" positions along the slit. The subtraction of the two consecutive exposures then allows an optimal removal of sky emission and instrumental background. The total exposure times are reported in Table \ref{tab:Logbook}.

GIANO-A, the precursor of GIANO-B, was fed by two fibres with a diameter on-sky of $1\arcsec$ and placed at a fixed projected distance of 3 arcsec. Similarly to GIANO-B, GIANO-A spectra were acquired following the nodding-on-fiber strategy by alternately observing the target through one fibre while the other fiber observed the sky.

\subsection{Ancillary observations}
To achieve an optimal flux calibration of the HARPS-N spectra, we obtained low-resolution optical spectra of our targets. The spectra ($\rm R=2400$, $\rm 330-790$ $\rm nm$) were acquired with the $\rm 1.22$ $\rm m$ telescope of the University of Padova, operated in Asiago (Italy). Observations were made in 2017 (October 27 and November 30), 2018 (December 6), 2019 (November 5-13), and 2020 (January 29 and February 6), which dates are close in time to the GIARPS observations. Spectra were reduced and flux-calibrated against a spectrophotometric standard. For most of the objects, the flux zero-point of the Asiago spectra was refined using the $\rm BVR_CI_C$ flux measurements (Table \ref{tab:Logbook}), collected with the ANS Collaboration telescopes \citep{Munari2012}.

Near-infrared photometric $JHK$ measurements were acquired in 2015 (December 17) using the SWIRCAM camera mounted on the Campo Imperatore $\rm 1.1$ $\rm m$ AZT-24 telescope \citep{Dalessio2000}, in 2017 (November 11), 2018 (December 22, 23, and 31), and 2019 (January 1, 10, and 11) with the REMIR instrument on the REM telescope \citep{Vitali2003}, located at the La Silla Observatory (Chile), and in 2020 (January 27) with the NICS camera \citep{Baffa2001} at the TNG telescope. The photometry of the objects is reported in Table \ref{tab:Logbook}.

\begin{table*}
\small
\center
\caption{\label{tab:Logbook} Journal of observations and photometry.}
\begin{tabular}{lcccccccccc}
\hline
\hline
          &              & $t_{exp}$ & $t_{exp}$ &       &       &       &       &       &       &      \\
Source    & Obs Date     & HARPS     & GIANO     & $B$   & $V$   & $R$   & $I$   & $J$   & $H$   & $K$  \\
\hline
          &              & [s]       & [s]       & [mag] & [mag] & [mag] & [mag] & [mag] & [mag] & [mag] \\
\hline
\hline   
 BP Tau          & 26 Gen 2020  & 3000		 & 2400		 & -	 & -	 & - 	 & -	 & 9.14  & -	& 7.81 \\
 CI Tau          & 09 Dec 2018  & 3000      & 2400      & 14.30 & 12.94 & 11.97 & 10.97 & 9.45  & 8.44 & 7.86 \\
 CoKu HP Tau G2  & 25 Gen 2020  & 3000		 & 2400		 & -	 & -	 & -	 & -	 & 8.18  & -	& 7.35 \\
 CQ Tau          & 13 Nov 2017  & 2200      & 1800      & 10.89 & 9.84  & 9.41  & 8.98  & 8.01  & 7.20 & 6.36 \\
 CW Tau          & 20 Dec 2015  & -         & 2400      & -     & -     & -     & -     & 9.10  & 7.82 & 6.74 \\ 
                 & 08 Dec 2018  & 3000      & 2400      & 15.35 & 13.76 & 12.53 & 11.51 & 9.05  & 7.80 & 6.95 \\    
 DD Tau          & 19 Dec 2015  & -         & 2400      & -     & -     & -     & -     & 10.10 & 8.78 & 7.86 \\ 
 DF Tau          & 20 Dec 2015  & -         & 1800      & -     & -     & -     & -     & 8.02  & 7.14 & 6.62 \\
                 & 08 Dec 2018  & 2200      & 1800      & 13.40 & 12.17 & 11.10 & 9.85  & 8.08  & 7.15 & 6.97 \\
 DG Tau          & 19 Dec 2015  & -         & 1800      & -     & -     & -     & -     & 8.51  & 7.30 & 6.36 \\
                 & 29 Oct 2017  & 2200      & 1800      & 13.65 & 12.58 & 11.50 & 10.46 & -     & 7.66 & 6.80 \\
 DH Tau          & 02 Nov 2019  & 3000      & 2400      & -     & -     & -     & -     & 9.48  & -    & 7.92 \\
 DK Tau          & 08 Dec 2018  & 2200      & 1800      & 13.96 & 12.58 & 11.59 & 10.59 & 8.89  & 7.90 & 7.04 \\
 DL Tau          & 20 Dec 2015  & -         & 2400      & -     & -     & -     & -     & 9.46  & 8.46 & 7.78 \\
                 & 29 Oct 2017  & 3000      & 2400      & 14.32 & 13.06 & 12.05 & 11.01 & 9.55  & 8.61 & 7.93 \\
 DN Tau          & 01 Nov 2019  & 3000      & 2400      & -     & -     & -     & -     & 9.13  & -    & 7.90 \\
 DO Tau          & 19 Dec 2015  & -         & 2400      & -     & -     & -     & -     & 9.89  & 8.61 & 7.57 \\
                 & 13 Nov 2017  & 3000      & 2400      & 14.34 & 13.18 & 12.27 & 11.23 & 9.28  & 8.17 & 7.34 \\ 
                 & 26 Gen 2020  & 3000      & 2400      & -     & -     & -     & -     & 9.29  & -    & 7.15 \\       
 DQ Tau          & 02 Nov 2019  & 3000      & 2400      & -     & -     & -     & -     & 9.52  & -    & 8.01 \\
 DR Tau          & 25 Gen 2020  & 2300		 & 1800		 & -	 & -	 & -	 & -	 & 8.89	 & -	& 6.80 \\
 DS Tau	         & 01 Nov 2019  & 3000      & 2400      & -     & -     & -     & -     & 10.31 & -    & 8.86 \\
 FT Tau          & 25 Gen 2020  & 4600		 & 3600		 & -	 & -	 & -	 & -	 & 10.29 & -	& 8.81 \\        
 GG Tau          & 09 Dec 2018  & 2200      & 1800      & 13.55 & 12.18 & 11.16 & 10.22 & 8.62  & 7.91 & 7.31 \\  
 GH Tau          & 25 Gen 2020  & 3000		& 2400		& -	    & -	 & -	 & -	 & 9.08	 & -	& 7.72 \\
 GI Tau          & 26 Gen 2020  & 3000		& 2400		& -	    & -	 & -	 & -	 & 9.56  & -    & 7.75 \\ 
 GK Tau          & 02 Nov 2019  & 3000      & 2400      & -     & -     & -     & -     & 8.89  & -    & 7.23 \\
 GM Aur          & 09 Dec 2018  & 3000      & 2400      & 13.41 & 12.30 & 11.44 & 10.71 & 9.49  & 8.79 & 8.61 \\
 HL Tau          & 24 Nov 2016  & -         & 1800      & -     & -     & -     & -     & -     & -    & -    \\
 HN Tau          & 19 Dec 2015  & -         & 3600      & -     & -     & -     & -     & 10.70 & 9.49 & 8.45 \\
                 & 29 Oct 2017  & 4500      & 3600      & 15.00 & 13.99 & 13.12 & 12.27 & 10.82 & 9.79 & 8.78 \\
 HQ Tau          & 01 Nov 2019  & 2300      & 1800      & -     & -     & -     & -     & 8.65  & -    & 7.13 \\
 IP Tau          & 09 Dec 2018  & 3000      & 2400      & 14.35 & 13.14 & 12.22 & 11.21 & 9.76  & 8.88 & 8.39 \\
 IQ Tau          & 02 Nov 2019  & 3000      & 3600      & -     & -     & -     & -     & -     & -    & 8.00 \\
 MWC480          & 01 Nov 2019  & 960       & 1200      & -     & -     & -     & -     & 7.82  & -    & 6.59 \\
 RW Aur A        & 13 Nov 2017  & 2200      & 1800      & 11.04 & 10.44 & 9.97  & 9.38  & 8.41  & 7.66 & 7.06 \\
 RY Tau          & 20 Dec 2015  & -         & 1200      & -     & -     & -     & -     & 7.40  & 6.30 & 5.40 \\
                 & 13 Nov 2017  & 1500      & 1200      & 11.49 & 10.36 & 9.61  & 8.81  & 6.95  & 6.45 & 5.84 \\
 UX Tau          & 26 Gen 2020  & 2280		& 1800		& -	 & -	 & -	 & -	 & 9.08  & - 	& 7.78 \\
 UY Aur          & 08 Dec 2018  & 2200      & 1800      & 13.99 & 12.72 & 11.71 & 10.78 & 9.15  & 8.07 & 7.15 \\
 UZ Tau E        & 09 Dec 2018  & 3755      & 3000      & 14.31 & 12.87 & 11.70 & 10.34 & 9.33  & 8.41 & 7.98 \\
 V409 Tau        & 26 Gen 2020  & 4600		& 3600		& -	 & -	 & -	 & -	 & 9.79  & -	& 8.44 \\
 V836 Tau        & 02 Nov 2019  & 3000      & 2400      & -     & -     & -     & -     & 9.91  & -    & 8.59 \\
 XZ Tau          & 14 Mar 2017  & 1800      & 3000      & -     & -     & -     & -     & -     & -    & -    \\
\hline\end{tabular}
\begin{quotation}
\textbf{Notes.} Typical errors in photometric magnitudes are 0.01 mag in the optical bands, 0.05 in the NIR bands for data taken in 2015,2017,2018 and 0.1 mag in the NIR bands for data taken in 2019-2020.
\end{quotation}
\end{table*}

\begin{table*}
\small
\center
\caption{\label{tab:SourceProperties} Source properties.}
\begin{tabular}{lcccccccc}
\hline
Source    & RA         & Dec           & d$^{(a)}$   & SpT          & $A_v$          & $L_\star$      & $M_{\star}$   & $i_{disk}^{(b)}$ \\
          & [J2000]    & [J2000]       & [pc]        &              & [mag]          & [$L_{\odot}$]  & [$M_{\odot}$] & [deg]            \\
\hline
\hline   
 BP Tau         & 04:19:15.83 & +29:06:26.93 & 129          & M0.5$^{(1)}$ & 0.45$^{(1)}$   & 0.42$^{(1)}$   & 0.52$^{(2)}$  & 38.2$^{(2)}$    \\
 CI Tau         & 04:33:52.01 & +22:50:30.09 & 158 		  & K5.5$^{(1)}$ & 1.90$^{(1)}$   & 0.63$^{(1)}$   & 0.90$^{(3)}$  & 50.0$^{(4)}$    \\
 CoKu HP Tau G2 & 04:35:54.15 & +22:54:13.40 & 166          & G2$^{(1)}$   & 2.55$^{(1)}$   & 6.92$^{(1)}$   & ...           & 50$^{(5)}$      \\ 
 CQ Tau         & 05:35:58.47 & +24:44:54.09 & 163 		  & F5$^{(6)}$   & 1.89$^{(7)}$   & 11.94$^{(7)}$  & 1.75$^{(7)}$  & 29$^{(8)}$      \\
 CW Tau         & 04:14:17.00 & +28:10:57.76 & 132 		  & K3.0$^{(1)}$ & 1.80$^{(1)}$   & 0.45$^{(1)}$   & 1.01$^{(3)}$  & 59$^{(9)}$      \\
 DD Tau         & 04:18:31.13 & +28:16:29.16 & 123 		  & M4.8$^{(1)}$ & 0.75$^{(1)}$   & 0.29$^{(1)}$   & 0.13$^{(1)}$  & ...             \\
 DF Tau         & 04:27:02.79 & +25:42:22.46 & 124 		  & M2.7$^{(1)}$ & 0.10$^{(1)}$   & 0.91$^{(1)}$   & 0.32$^{(3)}$  & 52$^{(3)}$      \\ 
 DG Tau         & 04:27:04.69 & +26:06:16.04 & 121 		  & K7.0$^{(1)}$ & 1.60$^{(1)}$   & 0.51$^{(1)}$   & 0.76$^{(3)}$  & 37$^{(9)}$      \\
 DH Tau         & 04:29:41.56 & +26:32:58.27 & 135 		  & M2.3$^{(1)}$ & 0.65$^{(1)}$   & 0.22$^{(1)}$   & 0.41$^{(3)}$  & 16.9$^{(2)}$    \\
 DK Tau         & 04:30:44.25 & +26:01:24.47 & 128 		  & K8.5$^{(1)}$ & 0.70$^{(1)}$   & 0.54$^{(1)}$   & 0.68$^{(3)}$  & 12.8$^{(2)}$    \\
 DL Tau         & 04:33:39.08 & +25:20:38.10 & 159 		  & K5.5$^{(1)}$ & 1.80$^{(1)}$   & 0.50$^{(1)}$   & 0.98$^{(2)}$  & 45.0$^{(2)}$   \\
 DN Tau         & 04:35:27.38 & +24:14:58.91 & 128 		  & M0.3$^{(1)}$ & 0.55$^{(1)}$   & 0.83$^{(1)}$   & 0.55$^{(3)}$  & 35.2$^{(4)}$    \\
 DO Tau         & 04:38:28.59 & +26:10:49.47 & 139 		  & M0.3$^{(1)}$ & 0.75$^{(1)}$   & 0.22$^{(1)}$   & 0.59$^{(2)}$  & 27.6$^{(2)}$   \\
 DQ Tau         &	04:46:53.06 & +17:00:00.13 & 197 		  & M0.6$^{(1)}$ & 1.40$^{(1)}$   & 1.17$^{(2)}$   & 1.61$^{(2)}$  & 16.1$^{(2)}$   \\
 DR Tau         & 04:47:06.21 & +16:58:42.81 & 196          & K6c$^{(1)}$  & 0.45$^{(1)}$   & 0.32$^{(1)}$   & 0.93$^{(2)}$  & 5.4$^{(2)}$     \\
 DS Tau	        & 04:47:48.59 & +29:25:11.19 & 159 		  & M0.4$^{(1)}$ & 0.25$^{(1)}$   & 0.19$^{(1)}$   & 0.69$^{(3)}$  & 65.2$^{(4)}$    \\
 FT Tau         & 04:23:39.19 & +24:56:14.11 & 128          & M2.8$^{(1)}$ & 1.30$^{(1)}$   & 0.18$^{(1)}$   & 0.34$^{(2)}$  & 35.5$^{(2)}$    \\
 GG Tau         & 04:32:30.33 & +17:31:40.83 & 140 		  & K7.5$^{(1)}$ & 1.00$^{(1)}$   & 1.41$^{(1)}$   & 0.62$^{(1)}$  & 57$^{(10)}$     \\
 GH Tau         & 04:33:06.22 & +24:09:33.63 & 130$^{(11)}$ & M2.3$^{(1)}$ & 0.40$^{(1)}$   & 0.64$^{(1)}$   & 0.36$^{(12)}$ & ...            \\
 GI Tau         & 04:33:34.06 & +24:21:17.07 & 130          & M0.4$^{(1)}$ & 2.05$^{(1)}$   & 0.56$^{(1)}$   & 0.52$^{(2)}$  & 43.8$^{(2)}$    \\
 GK Tau         &	04:33:34.56 & +24:21:05.85 & 129 		  & K6.5$^{(1)}$ & 1.50$^{(1)}$   & 0.93$^{(1)}$   & 0.69$^{(3)}$  & 40.2$^{(2)}$    \\
 GM Aur         & 04:55:10.98 & +30:21:59.37 & 159 		  & K6 $^{(3)}$  & 0.30$^{(1)}$   & 0.49$^{(1)}$   & 0.88$^{(3)}$  & 53.21$^{(13)}$ \\
 HL Tau         & 04:31:38.44 & +18:13:57.65 & 140 		  & K3c$^{(1)}$  & 2.50$^{(1)}$   & 0.14$^{(1)}$   & 1.8$^{(14)}$  & 47$^{(14)}$     \\
 HN Tau         & 04:33:39.36 & +17:51:52.30 & 145 		  & K3$^{(1)}$   & 1.15$^{(1)}$   & 0.17$^{(1)}$   & 1.53$^{(2)}$  & 69.8$^{(2)}$   \\
 HQ Tau         & 04:35:47.33 & +22:50:21.64 & 158 		  & K2.0$^{(1)}$ & 2.60$^{(1)}$   & 4.47$^{(1)}$   & 1.53$^{(3)}$  & 53.8$^{(2)}$    \\
 IP Tau         & 04:24:57.08 & +27:11:56.54 & 130 		  & M0.6$^{(1)}$ & 0.75$^{(1)}$   & 0.39$^{(1)}$   & 0.59$^{(3)}$  & 45.2$^{(4)}$    \\
 IQ Tau         & 04:29:51.56 & +26:06:44.85 & 131 		  & M1.1$^{(1)}$ & 0.85$^{(1)}$   & 0.24$^{(1)}$   & 0.50$^{(2)}$ & 62.1$^{(2)}$    \\
 MWC480         &	04:58:46.26 & +29:50:36.99 & 161 		  & A4.5$^{(4)}$ & 0.10$^{(2)}$  & 17.38$^{(1)}$  & 1.91$^{(2)}$ & 36.5$^{(2)}$   \\
 RW Aur A       &	05:07:49.54 & +30:24:05.07 & 163 		  & K0$^{(1)}$   & 0-2$^{(1,15)}$ & 0.72$^{(1)}$   & 1.20$^{(2)}$  & 55.1$^{(2)}$   \\	
 RY Tau         & 04:21:57.41 & +28:26:35.53 & 128 		  & F7$^{(4)}$   & 1.85$^{(1)}$   & 10.71$^{(1)}$  & 2.04$^{(2)}$  & 65.0$^{(2)}$   \\
 UX Tau         & 04:30:04.00 & +18:13:49.43 & 139$^{(11)}$ & M1.9$^{(1)}$ & 0.40$^{(1)}$   & 0.37$^{(1)}$   & 1.51$^{(3)}$  & 39.0$^{(12)}$   \\
 UY Aur         & 04:51:47.39 & +30:47:13.55 & 155 		  & K7.0$^{(1)}$ & 1.00$^{(1)}$   & 0.85$^{(1)}$   & 0.65$^{(2)}$ & 23.5$^{(2)}$   \\
 UZ Tau E       & 04:32:43.02 & +25:52:30.90 & 131 		  & M1.9$^{(1)}$ & 0.90$^{(1)}$   & 0.40$^{(1)}$   & 1.23$^{(2)}$  & 56.1$^{(2)}$   \\
 V409 Tau       & 04:18:10.78 & +25:19:57.38 & 131          & M0.6$^{(1)}$ & 1.00$^{(1)}$   & 0.66$^{(1)}$   & 0.50$^{(2)}$  & 69.3$^{(2)}$    \\
 V836 Tau       & 05:03:06.60 & +25:23:19.60 & 169 		  & M0.8$^{(1)}$ & 0.60$^{(1)}$   & 0.30$^{(1)}$   & 0.58$^{(3)}$  & 43.1$^{(2)}$   \\
 XZ Tau         & 04:31:40.09 & +18:13:56.64 & 144$^{(11)}$ & M2$^{(16)}$  & 1.40$^{(16)}$  & 0.42$^{(16)}$  & 0.36$^{(16)}$ & 35$^{(17)}$  \\
 \hline
\end{tabular}
\begin{quotation}
\textbf{Notes.} $^{(a)}$The distance for each target is computed from the \emph{Gaia} DR2 parallax \citep{Gaia2018}. $^{(b)}$ Disk inclinations are measured from ALMA observations and refer to the outer disk. An exception is made for GG Tau, where the inclination of the resolved inner disk  is reported.

\textbf{References.} $^{(1)}$\citet{Herczeg2014}, $^{(2)}$\citet{Long2019}, $^{(3)}$\citet{Simon2016}, $^{(4)}$\citet{Long2018}, $^{(5)}$\citet{Bouvier1995}, $^{(6)}$\citet{Varga2018}, $^{(7)}$\citet{Villebrun2019}, $^{(8)}$\citet{Chapillon2008}, $^{(9)}$\citet{Bacciotti2018}, $^{10}$\citet{Francis2020}, $^{(11)}$\citet{Akeson2019}, $^{(12)}$\citet{Banzatti2019}, $^{(13)}$\citet{Huang2020}, $^{(14)}$\citet{Yen2019}, $^{(15)}$\citet{Koutoulaki2019}, $^{(16)}$\citet{Hartigan2003}, $^{(17)}$\citet{Osorio2016}. 
\end{quotation}
\end{table*}

\section{Data reduction}
\subsection{HARPS/GIANO reduction}
The HARPS-N spectra were reduced using the standard HARPS-N Data Reduction Software  pipeline (\citealt{Pepe2002}), which includes the subtraction of the bias and dark current frames, the correction for the flat-field and scattered light, the wavelength calibration, and the 1D spectrum extraction. 
Spectra were corrected for the heliocentric and radial velocity using the profile of the \ion{Li}{I} photospheric line, assuming the weighted $\rm \lambda_{air}=670.7876$ $\rm nm$. 
The spectra of CQ Tau and MWC 480, which do not present the \ion{Li}{I} line, were calibrated using the photospheric lines of \ion{Fe}{II} at $\rm \lambda_{air}=450.8280$ $\rm nm$, and \ion{Si}{I} at $\rm \lambda_{air}=634.7109, 637.1371$ $\rm nm,$ respectively. The typical estimated wavelength accuracy of our calibration is about $1-2$ $\rm km$ $\rm s^{-1}$.

The \oi\ 630nm line profile was cleaned from telluric features through an interactive procedure and adopting as templates the standard stars acquired during the same observing night (see \citealt{Frasca2000}). Then, we used the tool \textsc{rotfit} \citep{Frasca2003} to fit and subtract the photospheric absorption lines of our targets with a grid of templates that were rotationally broadened. This procedure decreases the resolution of the final spectra, however, because the spectrum templates used within the \textsc{rotfit} tool have a spectral resolution of $R=42000$. For this reason, we used this procedure only on the targets for which the \oi\ profile is significantly altered by the photospheric features, that is, when photospheric lines are placed within the core of the \oi\ component with a depth of at least the 10\% of the local intensity.
For the objects whose \oi\ profile was not significantly contaminated, the cleaning was performed using the IRAF\footnote{IRAF is distributed by the National Optical Astronomy Observatory, which is operated by the Association of the Universities for Research in Astronomy, inc. (AURA) under cooperative agreement with the National Science Foundation.} task \textsc{splot} and applying a Gaussian line fitting. 

The GIANO-A spectra were reduced following the prescriptions given in \cite{Antoniucci2017}, while GIANO-B spectra were reduced according to the 2D GIANO-B data reduction prescriptions (see, e.g., \citealt{Carleo2018}).
Exposures with a halogen lamp were used to map the order geometry and for flat-field correction. The 1D spectra were obtained by extracting the spectral order containing the $\rm H_2$ profile with the use of the IRAF task \textsc{apall}. Finally, wavelength calibration was performed using a uranium-neon lamp acquired at the end of each night.  

The removal of the numerous telluric features present in the proximity of the $\rm H_2$ profile was performed in two steps. First, we created a synthetic telluric spectrum with the tool \textsc{molecfit} \citep{Smette2015}, which uses a radiative transfer code, a molecular line database, atmospheric profiles, and various kernels to model the instrumental line spread function. Second, we used the IRAF task \textsc{telluric} to correct the target spectra for the telluric contamination. This task divides the target spectrum by the synthetic telluric spectrum after the latter has been properly shifted and scaled to best match the observed telluric features.

Finally, the GIANO spectra were corrected for the heliocentric and radial velocity using three \ion{Al}{I} lines present in the same order as the $\rm H_2$ profile, that is, at $\rm \lambda_{vac}=2109.884, 2116.958, 2121.396$ $\rm nm$, with a typical final accuracy on the wavelength calibration of about $1-2$ $\rm km$ $\rm s^{-1}$.

\subsection{Flux calibration}
We flux-calibrated the HARPS-N data using the low-resolution spectra taken with the Asiago telescope, assuming that the continuum shape did not change significantly within the temporal distance between the HARPS-N and Asiago observations, which typically amount to $\sim 10$ days, except for two cases, where the Asiago observations were made after 30 days. Each HARPS-N spectrum was then continuum-normalized and multiplied for the curve resulting from a polynomial fit of the continuum of the Asiago spectrum.

The flux calibration of GIANO spectra was performed according to the procedures described in detail in \citet{Giannini2019}, that is, by multiplying the continuum-normalized GIANO order containing the $\rm H_2$ profile for a smooth continuum curve obtained from an interpolation of the $IJHK$ photometric points, after checking the good accordance between the $IJHK$ interpolation and the optical continuum Asiago spectrum.

\section{Results}

\begin{figure*}[!t]
\center
\includegraphics[trim= 600 440 60 60,width=0.7\columnwidth, angle=180]{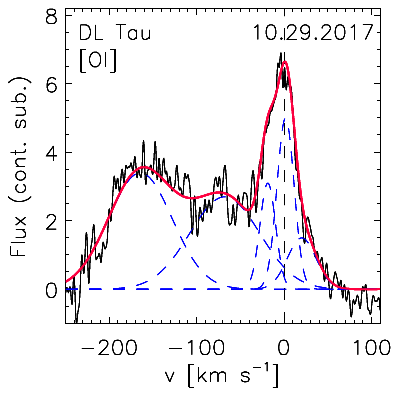}
\includegraphics[trim= 600 440 60 60,width=0.7\columnwidth, angle=180]{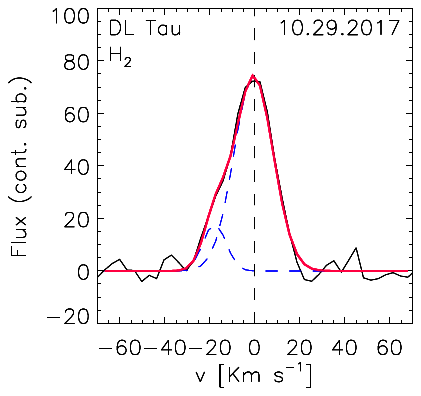}
\includegraphics[trim= 600 440 60 60,width=0.7\columnwidth, angle=180]{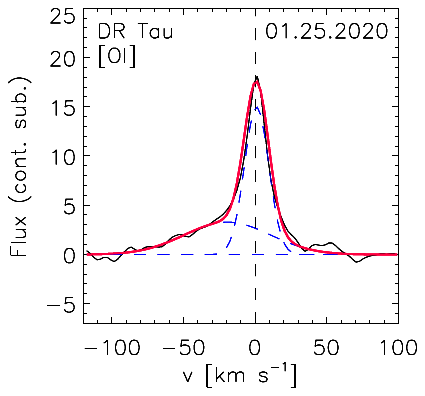}
\includegraphics[trim= 600 440 60 60,width=0.7\columnwidth, angle=180]{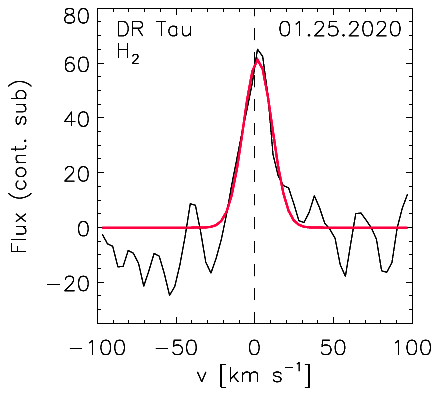}
\includegraphics[trim= 600 440 60 60,width=0.7\columnwidth, angle=180]{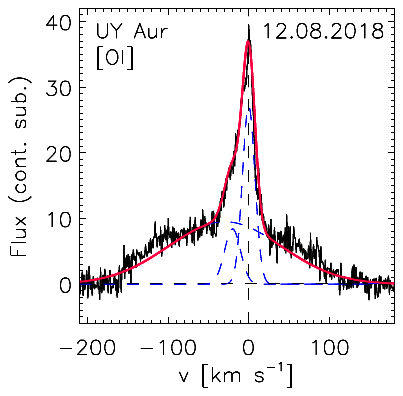}
\includegraphics[trim= 600 440 60 60,width=0.7\columnwidth, angle=180]{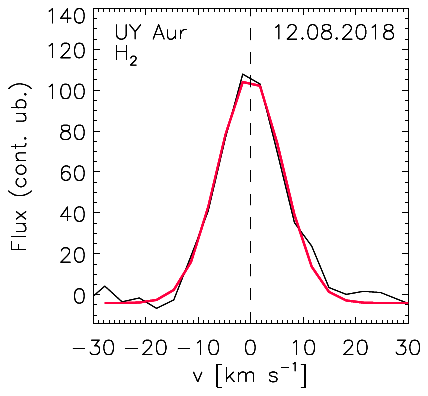}
\begin{center}\caption{\label{fig:OI_H2_profile_example} Example of Gaussian decomposition of continuum-subtracted \oi\ and \htwo\ line profiles (black lines). In red we plot the fit to the profile, and single components are shown as dashed blue lines. Flux units are $10^{-15}$ $\rm erg s^{-1}$ $\rm cm^{-2}$ $\AA^{-1}$. For each panel, target name and date of observation (MM.DD.YYYY) are indicated. The complete sample is reported in Fig. \ref{fig:OI_H2_profile}.}\end{center}
\end{figure*}

\begin{table*}
\small
\center
\caption{\label{tab:KinematicParameters} Line kinematic parameters of the NLVC for the \oi\ and \htwo\ profiles.}
\begin{tabular}{lc|ccc|ccc}
\hline
\hline
          &             & \multicolumn{3}{c}{NLVC - \oi}                     & \multicolumn{3}{c}{NLVC - \htwo}		                 \\	
Source    & Obs Date    & $\rm v_p$               & $\rm FWHM$              & $\rm I_p$ ($\times 10^{-15}$)                            & $\rm v_p$               & $\rm FWHM$              & $\rm I_p$ ($\times 10^{-15}$)               \\
          &             & [$\rm km$ $\rm s^{-1}$] & [$\rm km$ $\rm s^{-1}$] &  [$\rm erg$ $\rm s^{-1}$ $\rm cm^{-2}$ $\AA^{-1}$]  & [$\rm km$ $\rm s^{-1}$] & [$\rm km$ $\rm s^{-1}$] & [$\rm erg$ $\rm s^{-1}$ $\rm cm^{-2}$ $\AA^{-1}$]                       \\   
\hline   
 BP Tau         & 26 Gen 2020 & 2.5   $\pm$ 2.2  & 31.7 $\pm$ 4.4  & 16.4  $\pm$ 2.1  & 5.6    $\pm$ 2.4  & 27.3 $\pm$ 5.6   & 23.2 $\pm$ 3.1   \\  
 CI Tau         & 09 Dec 2018 & -1.5  $\pm$ 2.0  & 18.7 $\pm$ 2.0  & 2.7   $\pm$ 4.0  & ND                & ND               & ND               \\
 CoKu HP Tau G2 & 25 Gen 2020 & ND               & ND              & ND               & ND                & ND               & ND               \\  
 CQ Tau         & 13 Nov 2017 & 3.3   $\pm$ 3.4  & 29.3 $\pm$ 4.5  & 16.7  $\pm$ 2.0  & ND                & ND               & ND               \\    
 CW Tau         & 20 Dec 2015 & -                & -               & -                & ND                & ND               & ND               \\  
                & 08 Dec 2018 & 1.6   $\pm$ 4.2  & 35.2 $\pm$ 8.4  & 28.5  $\pm$ 3.0  & ND                & ND               & ND               \\ 
 DD Tau         & 19 Dec 2015 & -                & -               & -                & -19.1 $\pm$ 2.0   & 12.5 $\pm$ 0.7   & 42.0 $\pm$ 1.8   \\
 DF Tau         & 20 Dec 2015 & -                & -               & -                & ND                & ND               & ND                 \\     
                & 08 Dec 2018 & -7.8  $\pm$ 2.5  & 37.6 $\pm$ 4.7  & 9.8   $\pm$ 0.9  & ND                & ND               & ND                 \\
 DG Tau         & 19 Dec 2015 & -                & -               & -                & -17.1 $\pm$ 2.2   & 15.0 $\pm$ 2.1   & 235.0 $\pm$ 22.6 \\
                & 29 Oct 2017 & -6.8  $\pm$ 0.4  & 16.3 $\pm$ 0.2  & 205.0 $\pm$ 4.6  & -7.8  $\pm$ 1.0   & 14.2 $\pm$ 1.3   & 198.5 $\pm$ 12.7 \\
 DH Tau         & 02 Nov 2019 & 1.9   $\pm$ 0.4  & 44.5 $\pm$ 1.2  & 6.6   $\pm$ 0.1  & ND                & ND               & ND                 \\
 DK Tau         & 08 Dec 2018 & -2.4  $\pm$ 9.8  & 61.8 $\pm$ 18.5 & 18.0  $\pm$ 3.0  & ND                & ND               & ND                 \\     
 DL Tau         & 20 Dec 2015 &  -               & -               & -                &  2.6  $\pm$ 2.8   & 15.9 $\pm$ 4.7   & 103.5 $\pm$ 15.6 \\
                & 29 Oct 2017 &  2.7  $\pm$ 3.4  & 22.9 $\pm$ 5.9  & 5.0   $\pm$ 1.2  & -0.6  $\pm$ 1.4   & 19.1 $\pm$ 2.6   & 74.5 $\pm$ 8.6   \\
 DN Tau         & 01 Nov 2019 & -4.3  $\pm$ 2.0  & 45.2 $\pm$ 1.1  & 9.9   $\pm$ 0.2  & ND                & ND               & ND                 \\           
 DO Tau         & 19 Dec 2015 &  -               & -               & -                & -1.5  $\pm$ 2.4   & 9.2  $\pm$ 1.1   & 49.9 $\pm$ 6.0   \\
                & 13 Nov 2017 & -17.3 $\pm$ 2.7  & 34.6 $\pm$ 4.1  & 23.1  $\pm$ 1.8  & -17.3 $\pm$ 1.5   & 11.5 $\pm$ 2.7   & 60.0 $\pm$ 9.6   \\
                & 26 Gen 2020 & -14.3 $\pm$ 3.2  & 24.8 $\pm$ 7.7  & 24.5  $\pm$ 3.3  & -15.7 $\pm$ 2.1   & 11.1 $\pm$ 1.5   & 58.8 $\pm$ 6.0   \\  
 DQ Tau         & 02 Nov 2019 & -4.9  $\pm$ 2.1  & 26.9 $\pm$ 1.8  & 30.0  $\pm$ 0.9  & ND                & ND               & ND                 \\ 
 DR Tau         & 25 Gen 2020 & 0.9   $\pm$ 1.5  & 19.8 $\pm$ 1.4  & 15.0  $\pm$ 0.9  & 1.7   $\pm$ 1.    & 20.6 $\pm$ 2.3   & 61.3 $\pm$ 5.7   \\  
 DS Tau         & 01 Nov 2019 & 3.8   $\pm$ 2.2  & 52.5 $\pm$ 2.7  & 4.6   $\pm$ 0.2  &  1.4  $\pm$ 2.6   & 26.7 $\pm$ 4.1   & 3.7 $\pm$ 0.5    \\ 
 FT Tau         & 25 Gen 2020 & -18.0 $\pm$ 8.0  & 37.6 $\pm$ 10.6 & 1.3   $\pm$ 0.3  & ND                & ND               & ND                 \\  
 GG Tau         & 09 Dec 2018 & -0.4  $\pm$ 2.6  & 20.6 $\pm$ 5.8  & 10.5  $\pm$ 2.3  &  1.7  $\pm$ 4.0   & 22.5 $\pm$ 7.0   & 32.7 $\pm$ 4.8   \\
 GH Tau         & 25 Gen 2020 & -5.0  $\pm$ 4.3  & 68.2 $\pm$ 12.0 & 5.9   $\pm$ 0.7  & ND                & ND               & ND                 \\  
 GI Tau         & 26 Gen 2020 & -1.5  $\pm$ 3.7  & 28.1 $\pm$ 8.4  & 2.8   $\pm$ 0.5  & 1.8   $\pm$ 1.4   & 29.9 $\pm$ 3.3   & 50.6 $\pm$ 4.9   \\   
 GK Tau         & 02 Nov 2019 & -5.4  $\pm$ 2.0  & 36.6 $\pm$ 0.9  & 11.8  $\pm$ 0.2  & ND                & ND               & ND                 \\
 GM Aur         & 09 Dec 2018 & -0.9  $\pm$ 3.9  & 11.0 $\pm$ 4.9  & 17.8  $\pm$ 2.2  & -1.4  $\pm$ 1.7   & 17.6 $\pm$ 2.8   & 15.8 $\pm$ 2.2   \\ 
 HL Tau         & 24 Nov 2016 &  -               & -               & -                & -8.7  $\pm$ 3.0   & 12.0 $\pm$ 0.2   & -                  \\
 HN Tau         & 19 Dec 2015 &  -               & -               & -                & -1.5  $\pm$ 2.2   & 42.2 $\pm$ 2.1   & 23.3 $\pm$ 1.0   \\
                & 29 Oct 2017 & -7.5  $\pm$ 1.8  & 47.6 $\pm$ 1.3  & 31.5  $\pm$ 0.6  & -5.3  $\pm$ 3.4   & 37.6 $\pm$ 1.8   & 18.6 $\pm$ 1.0   \\ 
 HQ Tau         & 01 Nov 2019 & -6.4  $\pm$ 2.1  & 48.8 $\pm$ 1.9  & 7.4   $\pm$ 0.2  & ND                & ND               & ND                 \\ 
 IP Tau         & 09 Dec 2018 & -36.7 $\pm$ 2.4  & 43.9 $\pm$ 1.8  & 2.6   $\pm$ 0.7  & ND                & ND               & ND                 \\  
 IQ Tau         & 02 Nov 2019 & 1.8   $\pm$ 4.9  & 25.4 $\pm$ 13.4 & 21.5  $\pm$ 5.6  &  3.3  $\pm$ 2.5   & 15.0 $\pm$ 3.0   & 27.6 $\pm$ 0.7   \\   
 MWC 480        & 01 Nov 2019 & -13.2 $\pm$ 2.2  & 30.5 $\pm$ 2.1  & 38.8  $\pm$ 2.2  & ND                & ND               & ND                 \\  
 RW Aur A       & 13 Nov 2017 &  ND              & ND              & ND               & -2.6  $\pm$ 3.0   & 16.6 $\pm$ 5.2   & 72.0 $\pm$ 18.6  \\ 
 RY Tau         & 20 Dec 2015 & -                & -               & -                & ND                & ND               & ND                 \\ 
                & 13 Nov 2017 & -10.4 $\pm$ 6.2  & 38.9 $\pm$ 11.8 & 42.4  $\pm$ 9.4  & ND                & ND               & ND                 \\
 UX Tau         & 26 Gen 2020 & -2.0  $\pm$ 0.5  & 27.7 $\pm$ 2.2  & 48.5  $\pm$ 2.4  & -6.0  $\pm$ 1.1   & 11.2 $\pm$ 2.7   & 22.6 $\pm$ 4.1   \\  
 UY Aur         & 08 Dec 2018 & -0.5  $\pm$ 1.9  & 18.2 $\pm$ 3.4  & 27.1  $\pm$ 3.7  & -0.3  $\pm$ 1.7   & 12.7 $\pm$ 0.5   & 107.1 $\pm$ 6.0  \\
 UZ Tau E       & 09 Dec 2018 & -0.1  $\pm$ 5.4  & 54.1 $\pm$ 9.4  & 50.2  $\pm$ 7.5  &  1.3  $\pm$ 2.8   & 41.3 $\pm$ 5.8   & 15.8 $\pm$ 1.9   \\
 V409 Tau       & 26 Gen 2020 & 21.0  $\pm$ 12.9 & 54.1 $\pm$ 42.9 & 1.2   $\pm$ 0.4  & ND                & ND               & ND                 \\  
 V836 Tau       & 02 Nov 2019 & -4.2  $\pm$ 2.0  & 42.8 $\pm$ 1.2  & 4.8   $\pm$ 0.1  & ND                & ND               & ND                 \\
 XZ Tau         & 14 Mar 2017 & -0.9  $\pm$ 1.9  & 13.1 $\pm$ 3.8  & -                & -5.8  $\pm$ 1.8   & 16.9 $\pm$ 0.5   & -			        \\	
\hline
\end{tabular}
\begin{quotation}
\textbf{Notes.} ND: profile not-detected, "-": spectrum not-acquired.
\end{quotation}
\end{table*}

\begin{table*}
\small
\center
\caption{\label{tab:H2_OI_lum} \htwo\ and \oi\ extinction-corrected luminosity of the NLV component. }
\begin{tabular}{lccc}
\hline
Source           & Obs Date          & \htwo\      					    & \oi$_{\rm NLV}$\	  					 \\
                 &                   & [$10^{-5} \rm L_{\odot}$] 	& [$10^{-5} \rm L_{\odot}$]  \\
\hline
\hline   
 BP Tau          & 25 Gen 2020       & 3.3 $\pm$ 0.4   & 0.9 $\pm$ 0.1  	  \\ 	 
 CI Tau          & 09 Dec 2018       & < 1.5           & 0.4 $\pm$ 0.1     \\
 CoKu HP Tau G2  & 25 Gen 2020       & < 2.1		    & -			 		  \\ 
 CQ Tau          & 13 Nov 2017       & < 5.6           & 4.0 $\pm$ 0.2  	  \\  
 CW Tau          & 20 Dec 2015       & < 2.1           & -          		  \\ 
                 & 08 Dec 2018       & < 1.7           & 5.1 $\pm$ 0.2     \\ 
 DD Tau          & 19 Dec 2015       & < 2.2           & -			 		  \\ 
 DF Tau          & 20 Dec 2015       & < 2.1           & -          		  \\ 
                 & 08 Dec 2018       & < 1.6           & 0.4 $\pm$ 0.1     \\ 
 DG Tau          & 20 Dec 2015       & 14.8 $\pm$ 0.7  & -  	 	 		  \\ 
                 & 29 Oct 2017       & 11.9 $\pm$ 0.6  & 12.1 $\pm$ 0.5	  \\     
 DH Tau          & 02 Nov 2019       & < 1.4           & 0.6 $\pm$ 0.1     \\ 
 DK Tau          & 08 Dec 2018       & < 1.5           & 2.2 $\pm$ 0.1	  \\  
 DL Tau          & 20 Dec 2015       & 11.1 $\pm$ 0.6  & -      	 		  \\
                 & 29 Oct 2017       & 9.3  $\pm$ 0.5  & 0.8 $\pm$ 0.1	  \\
 DN Tau          & 01 Nov 2019       & < 1.2           & 0.8 $\pm$ 0.1 	  \\ 
 DO Tau          & 19 Dec 2015       & 2.6 $\pm$ 0.1   & -		     		  \\ 
                 & 13 Nov 2017       & 3.2 $\pm$ 0.4   & 1.9 $\pm$ 0.1	  \\ 
                 & 26 Gen 2020       & 3.5 $\pm$ 0.4   & 1.5 $\pm$ 0.1     \\     
 DQ Tau          & 02 Nov 2019       & < 1.5           & 6.6 $\pm$ 0.6 	  \\ 
 DR Tau          & 25 Gen 2020       & 12.5 $\pm$ 2.0  & 1.1 $\pm$ 0.1	  \\
 DS Tau	         & 01 Nov 2019       & 0.6  $\pm$ 0.1  & 0.5 $\pm$ 0.2     \\                  
 FT Tau          & 25 Gen 2020       & < 0.9    	   & 0.2 $\pm$ 0.1	  \\ 
 GG Tau          & 09 Dec 2018       & 3.7 $\pm$ 0.5   & 0.6 $\pm$ 0.1	  \\ 
 GH Tau          & 25 Gen 2020       & < 1.2		   & 0.6 $\pm$ 0.1	  \\ 
 GI Tau          & 26 Gen 2020       & 7.0  $\pm$ 0.7  & 0.5 $\pm$ 0.1	  \\ 
 GK Tau          & 02 Nov 2019       & < 1.5           & 1.6 $\pm$ 0.2	  \\ 
 GM Aur          & 09 Dec 2018       & 1.8 $\pm$ 0.2   & 0.4 $\pm$ 0.1	  \\ 
 HL Tau          & 24 Nov 2016       & -               & -		     		  \\
 HN Tau          & 19 Dec 2015       & 5.3 $\pm$ 0.2   & -     	 		  \\
                 & 29 Oct 2017       & 3.9 $\pm$ 0.2   & 4.9 $\pm$ 0.2	  \\
 HQ Tau          & 01 Nov 2019       & < 2.3           & 4.9 $\pm$ 0.5 	  \\                  
 IP Tau          & 09 Dec 2018       & < 1.3           & 0.2 $\pm$ 0.1     \\ 
 IQ Tau          & 02 Nov 2019       & 1.7 $\pm$ 0.2   & 1.3 $\pm$ 0.1	  \\ 
 MWC480          & 01 Nov 2019       & < 2.3           & 2.3 $\pm$ 0.8	  \\ 
 RW Aur A        & 13 Nov 2017       & 8.3 $\pm$ 0.7 - 6.4 $\pm$ 0.7$^{(1)}$     & - \\
 RY Tau          & 20 Dec 2015       & < 7.0           & -      	 		  \\ 
                 & 13 Nov 2017       & < 7.1           & 8.1 $\pm$ 0.4  	  \\ 
 UX Tau          & 26 Gen 2020       & 1.2 $\pm$ 0.4   & 2.5 $\pm$ 0.2	  \\ 
 UY Aur          & 08 Dec 2018       & 7.8 $\pm$ 0.5   & 1.8 $\pm$ 0.1     \\  
 UZ Tau E        & 09 Dec 2018       & 3.2 $\pm$ 0.2   & 6.6 $\pm$ 0.3	  \\ 
 V409 Tau        & 26 Gen 2020       & < 0.6		   & 0.2 $\pm$ 0.1 	  \\ 
 V836 Tau        & 02 Nov 2019       & < 1.2           & 0.7 $\pm$ 0.1	  \\		
 XZ Tau          & 14 Mar 2017       & -     		   & -			 		  \\
\hline
\end{tabular}
\begin{quotation}
\textbf{Notes.} (1) Different values are obtained for the adoption of different extinctions values.
\end{quotation}
\end{table*}

\subsection{Gaussian decomposition of the observed line profiles}
Figure \ref{fig:OI_H2_profile_example} shows examples of the observed \oi\ 630 nm and \htwo\ 2.12 $\rm \mu$m line profiles, while the complete sample is reported in Fig. \ref{fig:OI_H2_profile} of the Appendix. 
The \oi\ 630 nm line was detected in all sources except for CoKu HP Tau G2 and presents the typical composite profile, where components at different velocities can be identified. 

Following previous studies, we decomposed the \oi\ profiles into the different kinematic components by means of Gaussian fits. To do this, we developed a multicomponent line-fitting IDL procedure based on $\chi^2$ minimization. For each component, the procedure provides width, peak velocity, and peak intensity values. Errors were statistically estimated as the variation in the parameters that increases the $\chi^2$ of a unit. Following \cite{Banzatti2019}, the total number of components was determined as the minimum number of Gaussians that yields a $\chi^2$ stable at $\rm 20\%$ of its minimum value. In the case of XZ Tau and HN Tau, however, it was not possible to clearly distinguish between different components because the shape of the profiles is ramp-like, and the fitting procedure fails to converge. For these cases, we limited ourselves to identifying only the narrow LVC component by measuring the peak intensity in the |v$\rm_p|<30$ \kms\ range and fitting a single Gaussian basing on the red wing of the profile. 

We found that 20 sources (i.e., 60\% of the 34 sources for which we have the \oi\ spectrum) have a component at velocity $\rm |v\rm_p|> 30$ \kms\ (the HVC, associated with high-velocity jets), which in turn is sometimes fit by more than one Gaussian. The frequency of HVC detections in our sample is in line with that reported by \citet{Hartigan1995} and a factor two higher than the HVC detection rate reported by \citet{Nisini2018} in a sample of sources of the Lupus and Chameleon clouds. This result has been noted in \citet{Nisini2018} and was also ascribed, in addition to the lower resolution of the observations in this paper, to the on-average higher accretion rates of the sources in Taurus. 

Emission that peaks close to the systemic velocity (the LVC, |v$\rm_p| <$ 30 \kms) is detected in all sources except for RW Aur. This LVC was fit by a single Gaussian in 17 sources, and in 14 sources it was decomposed into two (three in a few cases) features. In the literature, the components identified at low velocity were usually separated into a narrow and a broad low-velocity component, that is, NLVC and BLVC, respectively, depending on their FWHM. We did not make this distinction because we sometimes found that what has previously been called a broad component is in fact due to the superposition of more than one Gaussian. The reason is that our spectral resolution
is higher than that of previous studies. 

In this paper we mostly discuss the properties of the narrow component at the lowest peak velocity (which is identical with the NLVC) as its properties are compared with the properties of the \htwo\ lines. We therefore report in Table\,\ref{tab:KinematicParameters} only its kinematic parameters, and the parameters of the other components we fit are given for completeness in Table\,\ref{tab:KinematicParametersAll} of the Appendix.

Comparing the parameters of our NLVC fit with those reported in the literature for the same sources (19 objects presented in \citealt{Simon2016,Banzatti2019}), we find that the peak velocity is consistent with the velocity reported in those works for the majority of cases, while the FWHM is different by up to a factor of two in ten cases, but without a recognizable trend. From an investigation on the origin of these discrepancies, we concluded that these differences are due to both an intrinsic variability of the line (e.g., DG Tau and DF Tau) and to the higher resolution (a factor of three) of our observations, which allows us to better define the separation among the different components.

The \htwo\ line is detected in 17 out of the  observed 36 sources (50\% detection frequency) and in most of the cases is single peaked and only slightly blueshifted (v$\rm_p <$ -20 \kms), consistent with previous \htwo\ observations of T Tauri stars at high resolution \citep[e.g.,][]{Beck2019}.
We applied the same Gaussian fitting procedure to the \htwo\ lines, although their profiles are not as structured as those of the \oi\ line. In the majority of cases, we identify a single Gaussian profile, with the exception of seven sources, for which a second Gaussian was needed to fit the excess emission in the blue wing. The fit parameters of the \htwo\ lines are reported in Tables \ref{tab:KinematicParameters} and \ref{tab:KinematicParametersAll}.
 
We also integrated the line flux of each kinematic component reproduced by the Gaussian fit. We report in Table \ref{tab:H2_OI_lum} the corresponding luminosity corrected for the extinction values reported in Table \ref{tab:SourceProperties} by assuming the extinction law by \citet{Cardelli1989}. In order to estimate the total uncertainty on the luminosity, we propagated the error on the line flux, which in turn was computed by considering the noise level in the proximity of the line and and the uncertainty on the flux calibration of the spectra.

\subsection{Photometric and spectroscopic variability}
\begin{figure*}[!t]
\center
\includegraphics[trim= 600 440 60 60,width=0.52\columnwidth, angle=180]{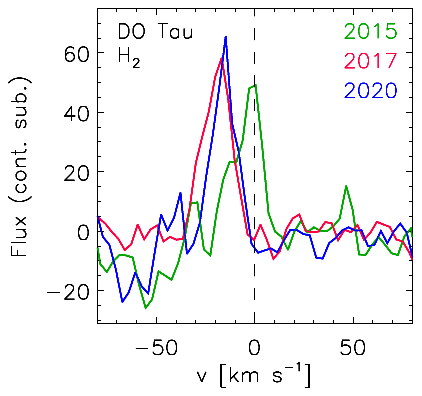}
\includegraphics[trim= 600 440 60 60,width=0.52\columnwidth, angle=180]{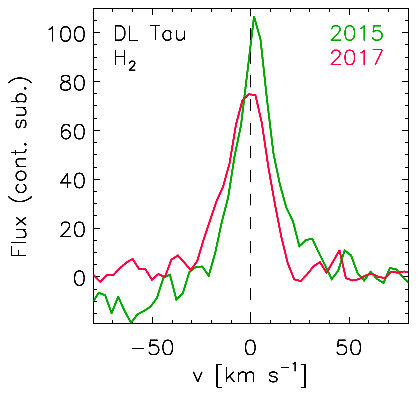}
\includegraphics[trim= 600 440 60 60,width=0.52\columnwidth, angle=180]{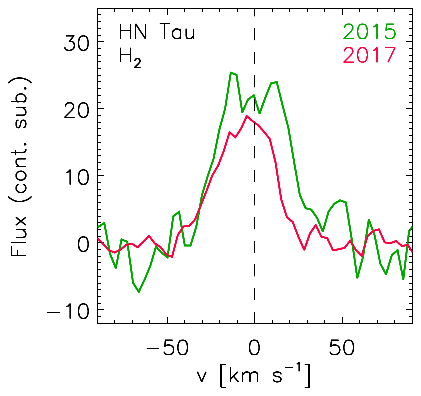}
\includegraphics[trim= 600 440 60 60,width=0.52\columnwidth, angle=180]{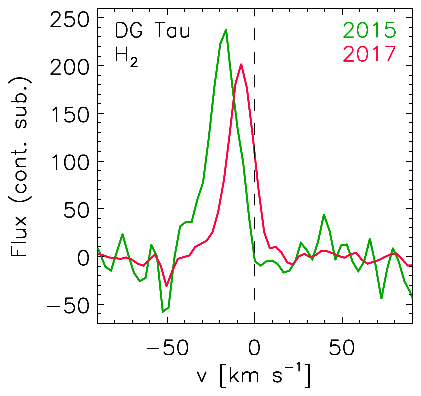}
\includegraphics[trim= 600 440 60 60,width=0.52\columnwidth, angle=180]{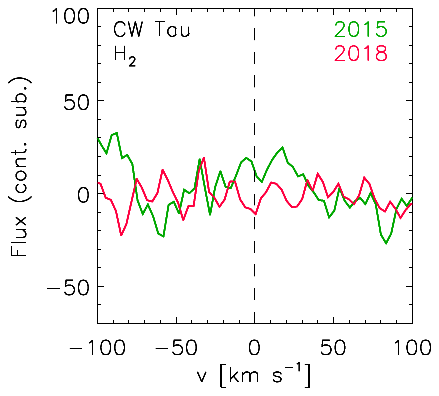}
\includegraphics[trim= 600 440 60 60,width=0.52\columnwidth, angle=180]{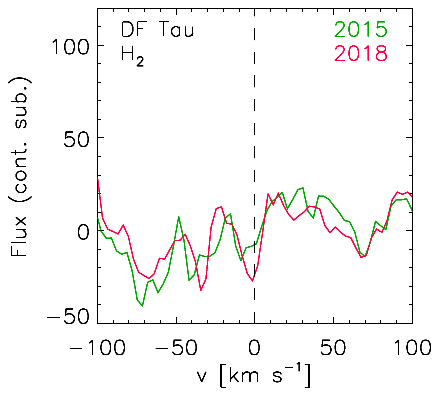}
\includegraphics[trim= 600 440 60 60,width=0.52\columnwidth, angle=180]{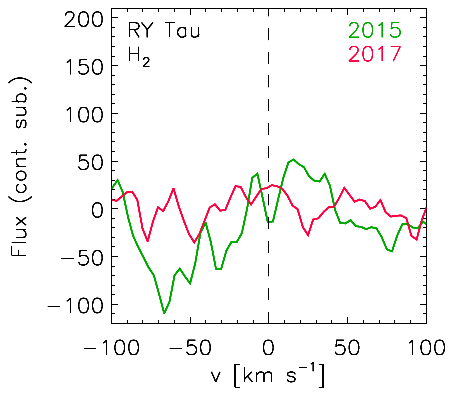}
\begin{center}\caption{\label{fig:TargetVariability} Continuum-subtracted \htwo\ line profiles variability for DO Tau, DL Tau, HN Tau, DG Tau, CW Tau, DF Tau and RY Tau. Flux units are $10^{-15}$ $\rm erg s^{-1}$ $\rm cm^{-2}$ $\AA^{-1}$.}\end{center}
\end{figure*}

As described in Sect. 2, seven sources were observed twice with the GIANO instrument. Ancillary photometry of these objects was taken on dates close to the spectroscopic observations and shows that the sources display photometric variations that range from a few tenths up to half a magnitude. The largest variations are observed in RY Tau and DO Tau and mostly affect the $J$ band, that is, the shortest wavelengths, as expected for young stellar objects that display accretion-related variability \citep[e.g.,][]{Lorenzetti2007}. 
The \htwo\ line profiles observed on different dates are displayed in Fig. \ref{fig:TargetVariability}. Relevant variations of the line profile, compatible with the resolution, is visible only in DO Tau and DG Tau. In particular, we observe a significant shift in the \htwo\ line peak of DO Tau, which was centered at zero velocity in 2015, while it appears at about $-15$ \kms\ in the 2017 and 2020 observations. Unfortunately, the 2015 observations did not include the HARPS-N spectrum, therefore we cannot evaluate whether the \oi\ 630 nm line experienced the same shift.

\subsection{Comparison between the \oi\ NLVC and the \htwo\ 2.12 \um\ emissions}
\begin{figure*}[!h]
\center\includegraphics[trim= 480 440 40 60,width=1.5\columnwidth, angle=180]{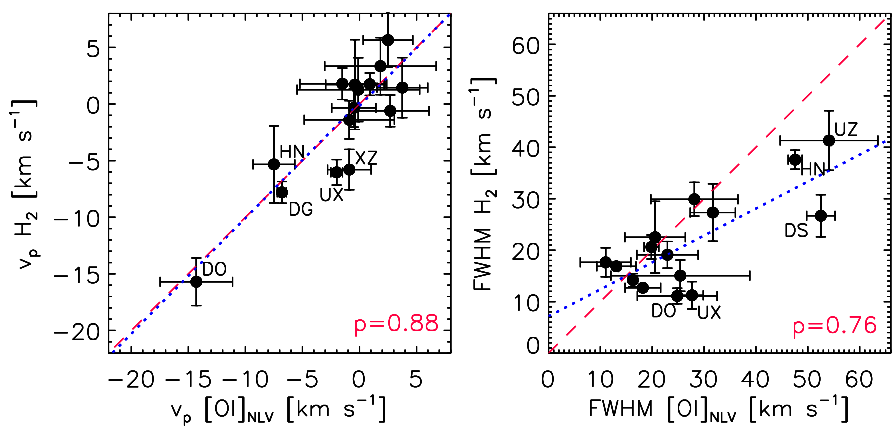}
\begin{center}\caption{\label{fig:Compare_OI_H2_fit} Correlations between kinematic properties of the \oi\ and \htwo\ NLV components. Left: Peak velocity; right: FWHM. The dashed red lines represent the one-to-one correlations, while linear fits are shown as dotted blue lines. The Pearson coefficient $\rm p$ is also reported. }\end{center}
\end{figure*}

The plots in Fig. \ref{fig:Compare_OI_H2_fit} show the peak velocity v$\rm_p$ and the FWHM of the \htwo\ line versus those of the \oi\ NLVC. The observed FWHM was deconvolved by the instrumental width $\rm \sigma_{instr}$ assuming a Gaussian profile for this latter ($\rm \sigma_{instr}=0.05$ $\AA$ for HARPSN and $\rm \sigma_{instr}=0.42$ $\AA$ for GIANO). We observe not only a correlation between the v$\rm_p$ of the two lines, but we also note that their values are remarkably similar within the uncertainties. In particular, five objects (DO Tau, DG Tau, HN Tau, UX Tau, and XZ Tau) display clearly blueshifted line peaks with velocities ranging between $\sim -$17 and $-$5 \kms. The peak velocities of all the other sources are around zero and are therefore compatible with quiescent emission or a very low wind velocity. The FWHMs of the two lines are also correlated, although the slope of the relations is not close to one, and in several cases, the FHWM of \htwo\ is smaller than that of \oi. In particular, we note that the width of the \htwo\ line is between 10 \kms\ and 40 \kms, while that of the \oi\, spans a wider range (10-60 \kms). 

\subsection{Dependence on disk inclinations}
\begin{figure*}[!h]
\begin{center}
\includegraphics[trim= 480 300 40 50,width=1.35\columnwidth, angle=180]{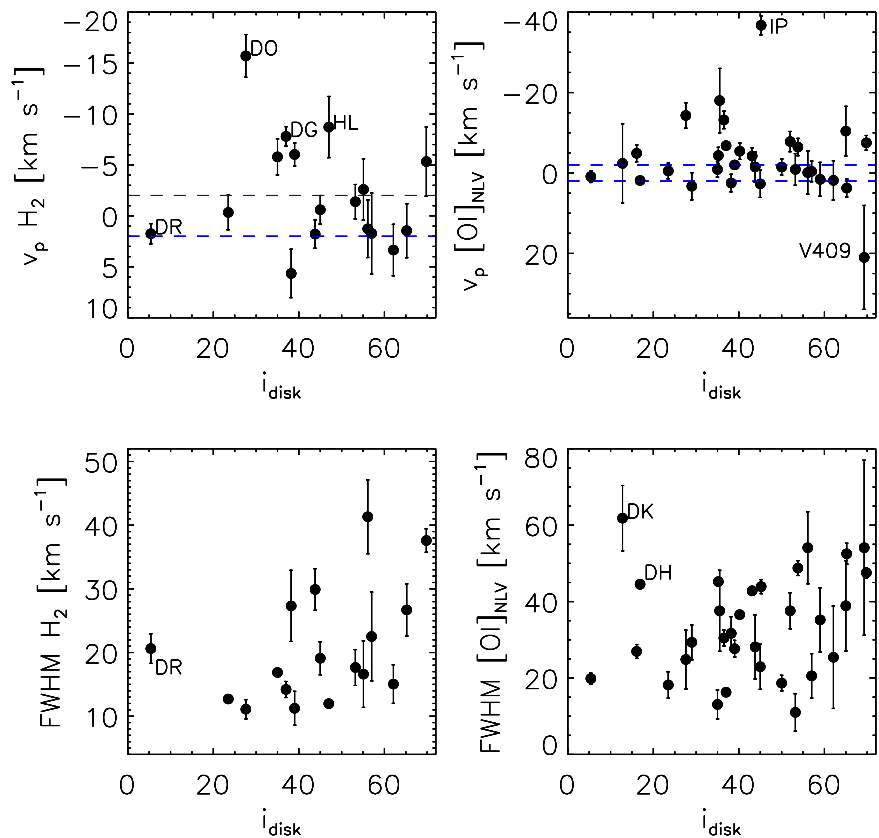}
\caption{\label{fig:Compare_incl_H2_fit}Correlations between disk inclination and kinematic properties of the \oi\ and \htwo\ NLVC. The dashed lbue lines represent the accuracy on the radial velocity calibration.}\end{center}
\end{figure*}

Figure \ref{fig:Compare_incl_H2_fit} (top) plots the peak velocity of the \htwo\ and \oi\ NLVC lines as a function of the disk inclination angle. A relation between these two quantities has been observed in the HVC of the \oi\ line for a sample of sources in Lupus, as expected if the HVC comes from collimated jets perpendicular to the disk \citep{Nisini2018}. Here, we do not see any clear trend in either line, as we expected because v$\rm_p$ of most stars is compatible with zero velocity. \citet{Banzatti2019} suggested that the broad, and to a lesser extent, also the narrow, LV components of the \oi\ line show the highest blueshift at inclinations of $\sim$ 35$\rm ^o$. We do not see this effect in our data. The reanalysis of the \citet{Banzatti2019} data performed by \citet{Weber2020} confirms our findings.  

Figure \ref{fig:Compare_incl_H2_fit} (bottom) shows the FWHM of the two lines versus the disk inclination angle $i\rm_{disk}$. Again, we do not see any clear correlation among these two quantities. A direct correlation between FWHM and sin$(i\rm_{disk})$ would be observed if the velocity width were caused by gas motion bound to the disk as in this case the velocity dispersion should be maximized for close to edge-on disks. However,
a correlation like this would only be observed if it were assumed that the line originates from the same disk region in all sources, while other parameters contribute in defining the region of emission. We return to this point in the next section. Here we just note that DR Tau appears to be an outlier in the FWHM(\htwo) versus sin$(i\rm_{disk})$ correlation. This source displays an almost face-on disk, whose inclination angle has recently been measured with ALMA observations to be about $\rm 5^o$  (\citealt{Long2019}). We mention, however, that interferometric IR observations of the inner disk of DR Tau are consistent with a higher inclination angle, and model fits to these observations suggest a value between 40$\rm ^o$ and 60$\rm ^o$ (Brunngr\"aber et al. 2016, Akeson et al. 2004). On these bases, the possibility that the inner and outer disks are misaligned should be taken into consideration. When we assume that the inner disk has an inclination of 40-60$\degr$, the position of DR Tau in the plots of Fig. \ref{fig:Compare_incl_H2_fit} would be more in line with the other sources of our sample. We should consider the possibility that this type of misalignment might indeed be common in young stellar objects. \citet{Francis2020} resolved the inner disk with ALMA in a sample of 14 transition disks and reported that about half of them are significantly misaligned with the outer disk. The only source in common with our sample is GG Tau, whose inclination angle is indeed about 20$\rm ^o$ higher than the outer disk. However, in sources with known resolved jets, the axis of the jet is always perpendicular to the disk inclination within few degrees \citep[e.g.,][]{Bacciotti2018}, which might indicate that for full disks, a misalignment like this is not so common. 
We also remark that a misalignment has also been observed between the outer disk inclination and the stellar rotation angle \citep[e.g.,][]{Davies2019}. This evidence does not imply, however, that the inner and outer disk are misaligned.

\section{Discussion}
We here discuss the link between the \htwo\ emission and the \oi\, emission in the NLVC. We determine whether this link holds any clue on the origin of the two line components that so far have been studied only individually.  

To help the discussion, Table\, \ref{tab:H2properties_v2} classifies the observed sources on the basis of the measured \htwo\ peak velocity.  |v$\rm_p| >4$ \kms\ indicates a clear origin in a wind, and |v$\rm_p| <4$ \kms\ means that a source is also compatible with gas bound in the disk. We assumed 4 \kms\ as our resolution limit, that is, twice the typical accuracy of the wavelength calibration. The table also summarizes the type of the observed \htwo\ and \oi\ profiles. Additional information retrieved from the literature relative to the spatial extent of the emission or to the inner disk properties is also indicated.

\begin{table*}
\small
\center
\caption{\label{tab:H2properties_v2} Emission properties.}
\begin{tabular}{lccclll}
\hline
\\
Source    & \htwo\ profile$^a$        & \oi\ profile$^a$  & R$_k$(H$_2$)$^b$  & R$_k$(\oi) $^b$ &  Other information & Ref.\\
          &                           &                   & [au]              & [au]             &                          \\
\hline
\hline
\\
\multicolumn{6}{c}{Sources with |v$\rm_p$(\htwo)| $>4$\,\kms} \\
\\
\hline   
 DD Tau  & S & ...   &                  & ...             & \oi\ HV jet                         & (1)      \\
 DG Tau  & D & D/HVC & 4.84 $\pm$ 0.86  & 3.69 $\pm$ 0.09 & \htwo\ extended, wind               & (2),(3)  \\  
 DO Tau  & D & S/HVC & 3.65 $\pm$ 0.99  & 0.73 $\pm$ 0.45 &                                     &          \\
 HN Tau  & S & D/HVC & 3.38 $\pm$ 0.33  & 2.11 $\pm$ 0.12 &                                     &          \\
 HL Tau  & S & ...   & 23.84 $\pm$ 1.00 & ...             & \oi\ HV jet,  \htwo\ extended, wind &  (4),(5) \\
 UX Tau  & S & S/HVC & 16.79 $\pm$ 7.95 & 2.77 $\pm$ 0.44 & TD ($R_{gap}$=25 au )               &          \\
 XZ Tau  & S & D/HVC & 1.48 $\pm$ 0.09  & 2.46 $\pm$ 1.42 & \htwo\ extended, wind               & (2)      \\  
\hline
\\
\multicolumn{6}{c}{Sources with |v$\rm_p$(\htwo)| $<4$\,\kms}\\
\\
\hline
 BP Tau  & S & S/HVC & 0.94 $\pm$ 0.39 & 0.70  $\pm$ 0.19  &                                     &           \\
 DL Tau  & D & D/HVC & 4.76 $\pm$ 1.29 & 3.30  $\pm$ 1.70  &                                     &           \\
 DR Tau  & S & S/HVC & 0.07 $\pm$ 0.02 & 0.07  $\pm$ 0.01  &                                     &           \\ 
 DS Tau  & S & S     & 2.83 $\pm$ 0.87 & 0.73  $\pm$ 0.08  &                                     &           \\
 GG Tau  & D & S/HVC & 1.43 $\pm$ 0.88 & 1.71  $\pm$ 0.97  &\htwo\ extended, disk/infall streams & (6)       \\ 
 GI Tau  & S & D/HVC & 0.99 $\pm$ 0.22 & 1.11  $\pm$ 0.67  &                                     &           \\
 GM Aur  & S & S     & 6.43 $\pm$ 2.05 & 16.51 $\pm$ 14.61 & \htwo\ compact, disk; TD ($R_{gap} \sim$ 30-40 au ) & (6), (7)  \\
 IQ Tau  & D & S/HVC & 6.12 $\pm$ 2.47 & 2.14  $\pm$ 2.26  &                                     &           \\
 UY Aur  & S & D/HVC & 2.27 $\pm$ 0.19 & 1.11  $\pm$ 0.42  & \htwo\ extended, wind, binary                       & (6)       \\
 UZ Tau E & S & S/HVC & 1.76 $\pm$ 0.49 & 1.03  $\pm$ 0.36  & TD ($R_{gap} \sim$ 4.7 au )                         & (8)       \\
\hline
\\
\multicolumn{6}{c}{Sources with no \htwo\ detections} \\
\\
 \hline
CI Tau     & ... & D     & ... & 5.38 $\pm$ 1.17 & \oi\ weak                      & \\
CoKu HP Tau G2      & ... & ...   & ... & ...             & no \oi\                        & \\
CQ Tau     & ... & D     & ... & 1.69 $\pm$ 0.53 & F-type star                    & \\
CW Tau     & ... & D/HVC & ... & 2.12 $\pm$ 1.01 &                                & \\
DF Tau     & ... & D/HVC & ... & 0.50 $\pm$ 0.12 & \oi\ at R<1                    & \\ 
DH Tau     & ... & S     & ... & 0.06 $\pm$ 0.01 & \oi\ at R<1                    & \\     
DK Tau     & ... & S     & ... & 0.03 $\pm$ 0.01 & \oi\ at R<1                    & \\
DN Tau     & ... & S     & ... & 0.32 $\pm$ 0.01 & \oi\ at R < 1                  & \\ 
DQ Tau     & ... & D/HVC & ... & 0.60 $\pm$ 0.08 & \oi\ at R <1                   & \\ 
GH Tau     & ... & S/HVC & ... & ...             & \oi\ at R <1  (no inc. angle)  & \\
GK Tau     & ... & S     & ... & 0.76 $\pm$ 0.04 & \oi\ at R <1                   & \\ 
HQ Tau     & ... & S     & ... & 1.48 $\pm$ 0.12 & \oi\ low S/N                   & \\
IP Tau     & ... & S     & ... & 0.55 $\pm$ 0.05 & weak OI  at R<1, TD ($R_{gap} \sim$ 21 au )      & (8) \\
MWC480     & ... & S     & ... & 2.58 $\pm$ 0.35 & A-type                         & \\ 
RW Aur A   & ... & D/HVC & ... & ...             & no \oi\ NLVC                   & \\
RY Tau     & ... & S/HVC & ... & 3.93 $\pm$ 2.38 & F-type, TD ($R_{gap} \sim$ 6.9 au ) & (8) \\
V409 Tau   & ... & D/HVC & ... & 0.53 $\pm$ 0.45 & \oi\ at R <1                   & \\
V836 Tau   & ... & S     & ... & 0.52 $\pm$ 0.03 & \oi\ at R <1                   & \\
\hline 
\hline \\
\hline
\end{tabular}
\begin{quotation}
\textbf{Notes.} $^a$ S: LVC with single component profile, D: LVC with double component profile (the other component might have V$_p > |3|$ \kms), HVC: presence of at least one high velocity component. The HVC is defined as a component at V$_p >$ 30 km/s, taking also into account corrections for inclination angles. $^b$ Estimated emitting radius for the emission assuming keplerian broadening.

\textbf{References.} (1) Hartigan et al. (2004), (2) \citet{Beck2008} , (3) \citet{Agra-Amboage2014}, (4) \citet{Mundt1990}, (5) \citet{Takami2007}, (6) \citet{Beck2019}, (7) \citet{Hornbeck2016}, (8) \citet{Long2018}.  \end{quotation}
\end{table*}

\subsection{Comparison with spatially resolved observations}
The small-scale spatial extent of the \htwo\ emission in T Tauri stars has been investigated in several studies by employing adaptive optics-assisted spectroimaging  (e.g., Beck \& Bary 2019, Agra-Amboage et al. 2014, Back et al. 2008, Takami et al. 2007). Of the seven sources that show a clear blueshifted \htwo\ line in Table\,\ref{tab:H2properties_v2}, imaging in \htwo\ is available for DG Tau, XZ Tau, and HL Tau. The extended emission in all of them is compatible with poorly collimated winds. Of the sources in which \htwo\ has |v$\rm_p| < 4$ \kms, UY Aur is the only one in which a wind-like extended \htwo\, emission has been observed. Conversely, imaging of GM Aur shows a resolved but compact emission compatible with an origin in the disk, while the \htwo\ extended emission from GG Tau has been interpreted as originating in infall streams from the outer circumbinary disk to the inner region. Therefore,  the little statistics we have at hand appears to suggest that \htwo\ is mostly excited in extended winds, and in particular, a wind origin is observed whenever the source also displays a \oi\ HVC indicative of jets (see Table\,\ref{tab:H2properties_v2}). Additional high angular resolution observations are needed to confirm this association on a larger statistical basis. 

Direct comparisons between spatially resolved images of \htwo\ and \oi\ NLVC are lacking at present because current facilities do not provide the high spatial and spectral resolution needed to spatially separate the \oi\ NLVC from components at higher velocity. We therefore cannot verify whether the derived kinematic link between the two emissions corresponds to a similar spatial extent. In DG Tau, where \oi\ spectral imaging in different velocity bins has reported in the literature \citep[e.g.,][]{Lavalley-Fouquet2000,Maurri2014}, the \oi\ LVC extends to larger distances than \htwo\ (i.e., $\sim 1\arcsec$ vs $0.5\arcsec$) while the spatial width of the two emissions is comparable (i.e., $0.2-0.3\arcsec$ at $0.2-0.5\arcsec$ from the source). However, the \oi\ LVC as defined in these works covers a wider velocity range of [$-100,10$] \kms\ that also encompasses the other LVC components and part of the HVC in our observed line profile. 

\subsection{Comparison with previous studies on the \oi\, emission}

The NLVC in the composite profile of the \oi\,630 nm line was originally reported by \cite{Rigliaco2013} based on the analysis of high-resolution spectra of the \oi\, line in two T Tauri stars. 
More recently, \cite{Simon2016} and \cite{Banzatti2019} have analyzed this emission on larger samples of T Tauri stars, which allowed them to define its properties based on a statistical analysis. These works highlight in particular the following properties: 1) the narrow and broad LVCs correlate with the accretion luminosity, but the NLVC is more often detected in low accretors. In particular, the NLVC has been found to be proportionally stronger in transition disk sources, that is, in sources showing a significant depletion of dust in their inner region. 2) The NLVC appears to be more blueshifted when a HVC (jet) is present as well, suggesting a link between these two emissions. 3) The NLVC FWHM correlates with the IR index, which is an indicator of the dust content in the inner disk, showing that the \oi\, emission recedes to larger radii for more evolved disks. 
Based on these pieces of evidence, it has been suggested that in sources with a dominant HVC, both broad and narrow LVC are part of the same MHD wind that feeds HV jets. In more evolved sources, where the action of magnetocentrifugal forces giving rise to high-velocity jets has already faded, the NLVC might be due to a residual low-velocity MHD wind, to photoevaporative winds, or even to gas bound in the disk. 
We now determine how the \htwo\ emission fits into this picture.

As shown in Table\,\ref{tab:H2properties_v2}, we find that sources with a clear blueshifted v$\rm_p$ in both \htwo\ and \oi\ are always associated with a HVC (or known collimated jets for DD Tau and HL Tau, for which we do not have \oi\ data). The reverse is not always true:   v$\rm_p$ in several of the objects
with a prominent HVC is compatible with no shift with respect to the source velocity, therefore we cannot confirm a strict kinematic association between  \oi\ NLVC, \htwo\, and the high-velocity jets in our sample. 

Of the sources without a HVC, only two have detected \htwo\ emission (GM Aur and DS Tau). As we have discussed in the previous section, the spatially compact \htwo\ emission of GM Aur is compatible with a disk origin. \htwo\ and \oi\ FWHM are similar in this source, which might suggest that the \oi\ emission could also come from the disk. In general, we note that the \htwo\ detection frequency is higher in sources with a HVC (15 out of 22 with respect to 2 out of 12). On the other hand, in sources in which no \htwo\ emission is detected, the \oi\ HVC was revealed in 7 out of 18. We conclude that while the emission in MHD winds might indeed be the prominent mechanism of the \htwo\ excitation, the HVC in the \oi\ emission does not necessarily imply that \htwo\ emission is detected.

Finally, we identified five transition disk sources in our sample (see Table 5): \htwo\ has been detected in three of them (UX Tau, GM Aur, and UZ Tau), and only one (GM Aur) has a single narrow component of the \oi\ emission. We therefore conclude that our statistics is too small to assess whether \htwo\ in TD sources behaves differently than in the other sources.

\subsection{\htwo\, luminosity correlations}

\begin{figure}[!t]
\includegraphics[trim= 480 300 40 190,width=1.7\columnwidth, angle=180]{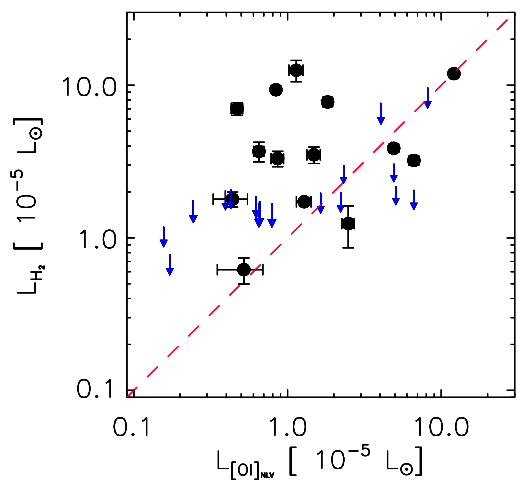}
\begin{center}\caption{\label{fig:Compare_OI_H2_lum} Correlations between extinction-corrected luminosities of the \oi\ and \htwo\ NLVC. The dashed red line represents the one-to-one correlation. Upper limits are indicated with blue arrows.}\end{center}
\end{figure}

Fig. \ref{fig:Compare_OI_H2_lum} plots the \htwo\, line luminosity as a function of the luminosity of the \oi\ NLVC. There is no clear trend between the two luminosities in the small luminosity range we probed.

Fig. \ref{fig:Compare_OI_H2_lum} shows that about half of the sources in which \htwo\, has been detected have a similar \htwo\ and \oi\ luminosity. In the other half of the sample, the \htwo\ luminosity is significantly higher than the \oi\ luminosity. The plot also shows that most of the upper limits on the \htwo\ luminosity are consistent with the median values of the detections.

\begin{figure*}[!h]
\includegraphics[trim= 480 300 40 60,width=0.9\columnwidth, angle=180]{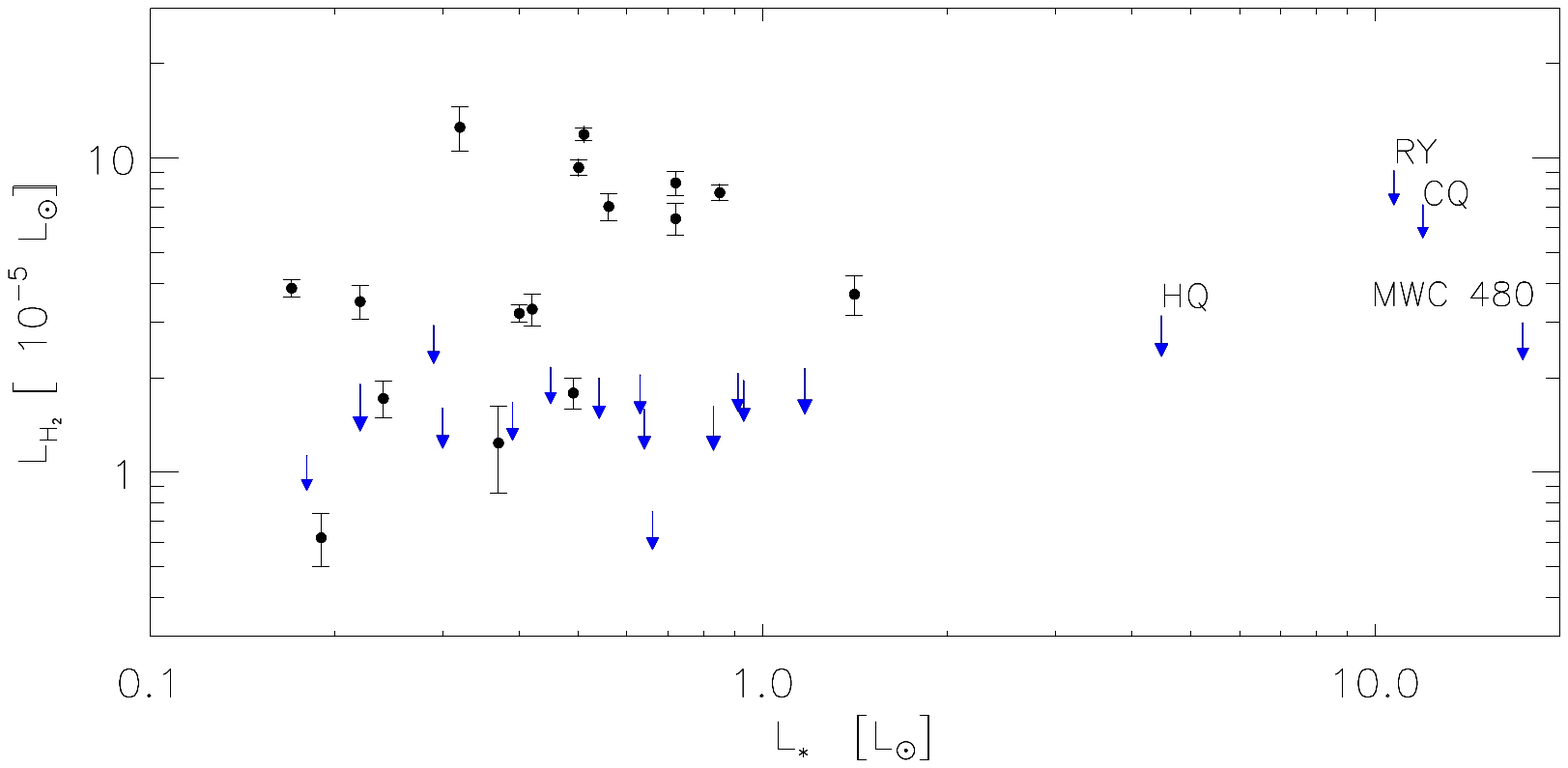}
\begin{center}\caption{\label{fig:lum_star_vs_lum_H2}Correlation between stellar luminosities and extinction-corrected luminosities of the \htwo\ NLVC. Upper limits are indicated with blue arrows.}\end{center}
\end{figure*}

\begin{figure*}[!h]
\includegraphics[trim= 480 300 40 60,width=0.9\columnwidth, angle=180]{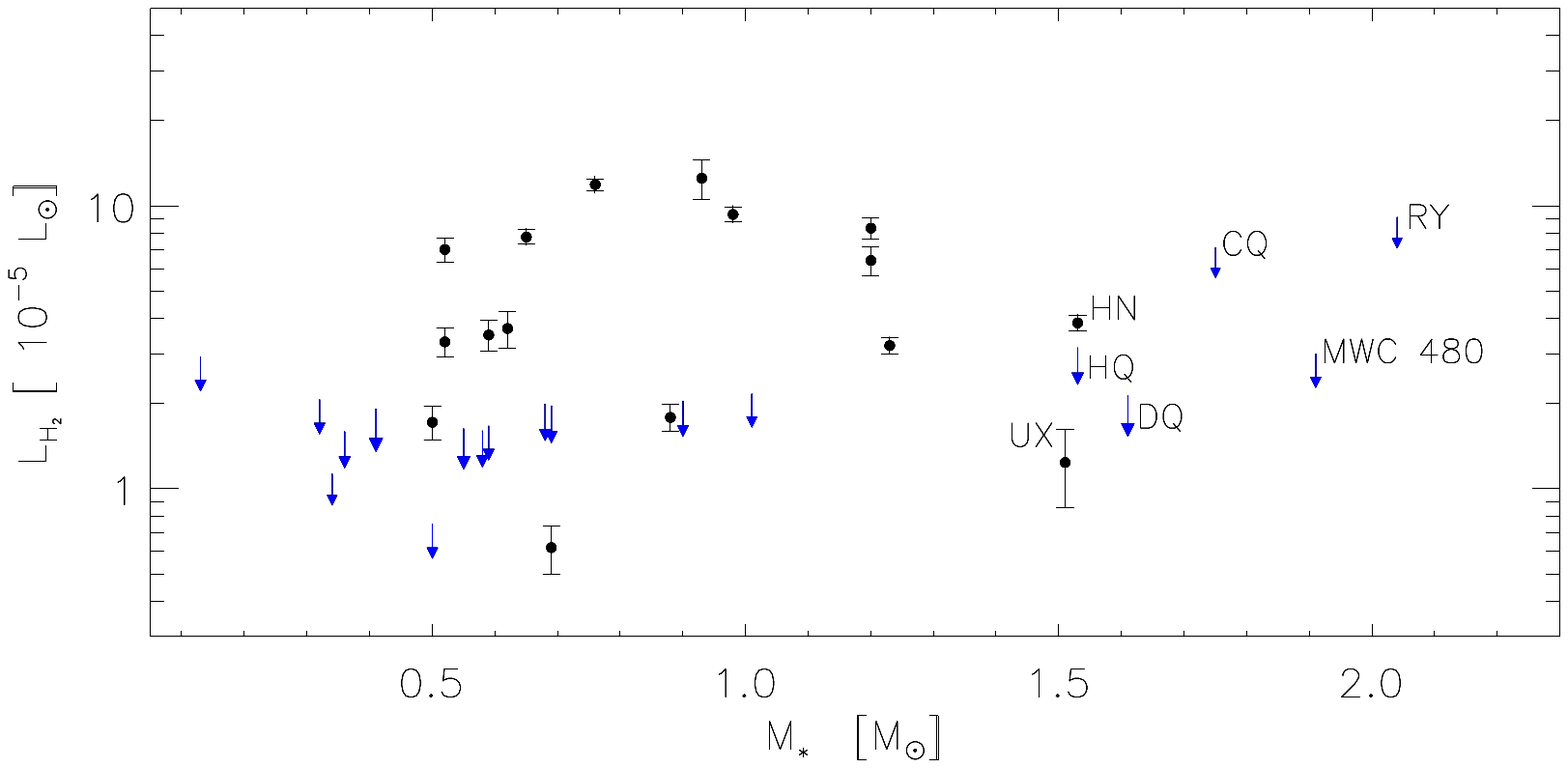}
\begin{center}\caption{\label{fig:mass_vs_lum_H2}Correlation between stellar masses and extinction-corrected luminosities of the \htwo\ NLVC. Upper limits are indicated with blue arrows.}\end{center}
\end{figure*}

To better clarify the dependence of the \htwo\ luminosity on the source properties, we plot in Figs.\,\ref{fig:lum_star_vs_lum_H2} and \ref{fig:mass_vs_lum_H2} the \htwo\ luminosity as a function of the stellar luminosity and mass. No clear trend is apparent in these plots. We note, however,  that \htwo\, is never detected in the most luminous (i.e., $\rm L_* >$ 2 L$_\odot$) and massive (i.e., $\rm M_* >$ 1.5 M$_\odot$) sources. This evidence might suggest that in luminous early-type sources, a stronger radiation field from the central star might dissociate \htwo\ in the disk surface layers and thus inhibit the formation of molecular winds.

We therefore conclude that the relative luminosity of \oi\ and \htwo\ lines does not have a clear recognizable trend and might be affected by a different degree of dissociation, ionization, or excitation conditions in the emitting region. We discuss the excitation conditions in the next section.

\begin{figure*}[!t]
\center\includegraphics[trim= 480 300 40 60,width=1.7\columnwidth, angle=180]{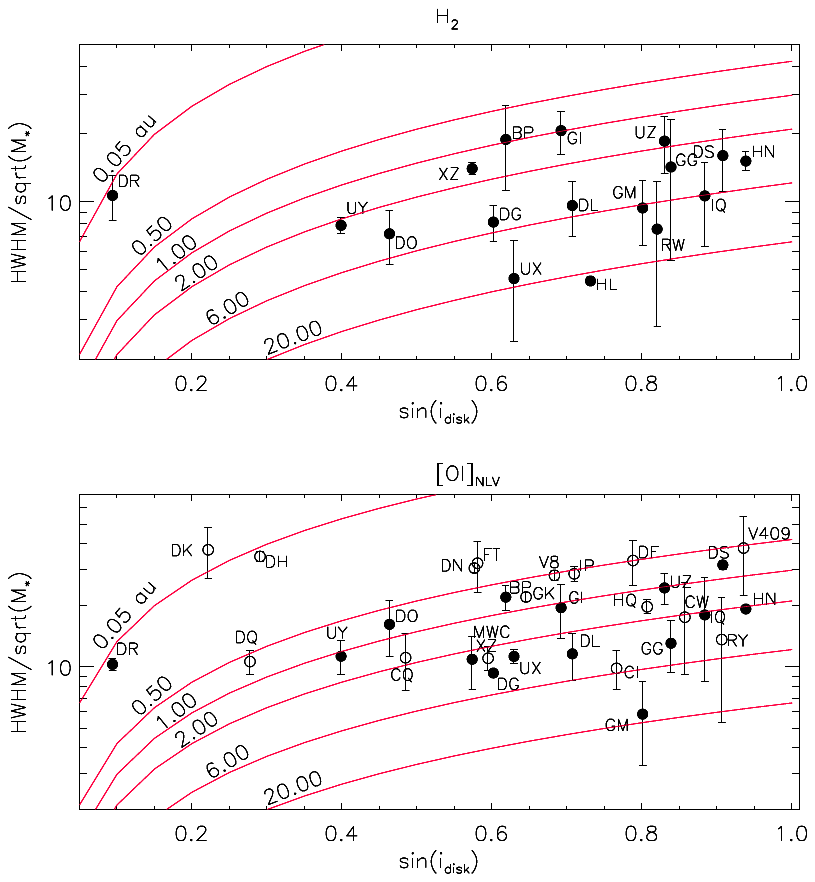}
\begin{center}\caption{\label{fig:Compare_incl_OI_fit}Half-width at half maximum divided by the square root of the stellar mass for the \htwo\ (top) and \oi\ (bottom) NLVC as a function of the sine of the disk inclination. Keplerian models for gas emitted from disk radii of 0.05, 0.50, 1.00, 2.00, 6.00, and 20.00 au are shown as solid red lines. Bottom panel: Empty circles refer to stars whose \htwo\ profile is not detected.}\end{center}
\end{figure*}

\subsection{Emission size of the  \htwo\, and NLVC \oi\ lines}

We can obtain insights into the size of the emission region by analyzing the observed line widths. When we assume that the line width is dominated by Keplerian broadening, we can derive the radius of the region of the disk where the bulk of the considered components originates from the relationship $\rm R_K = (sin(i\rm_{disk})/HWHM)^2\,\times G\,\times M_\star$, where $i\rm_{disk}$ is the disk inclination angle and HWHM is the line half-width at half maximum.

If the emission is due to a slow wind, this assumption remains valid as soon as the gas does no acquire a significant acceleration as it is lifted from the disk. This prescription has largely been applied in different studies of the \oi\ line \citep{Simon2016,Mcginnis2018,Banzatti2019}. However, recent models of photoevaporative and MHD winds (Weber et al. 2020) have explored the validity of Keplerian motion for forbidden atomic lines, showing that vertical velocity gradients might significantly contribute to the line broadening, especially at low inclination angles. In particular, caution should be applied when the $R_K$ values are interpreted as a measure of the size of the emitting region in sources with an inclination $\la 20$ degrees.

With this caveat in mind, we plot in Fig.\,\ref{fig:Compare_incl_OI_fit} the HWHM divided by the square root of the stellar mass as a function of $\rm sin(i\rm_{disk})$. In the plot, lines of constant Keplerian radius are drawn to guide the eye. 
The observed widths imply emitting radii between 0.05 and $\sim$20 au for \oi, and between 2 and 20 au for \htwo. The range of radii estimated for \oi\ is in line with what was found in other studies for the component considered here (e.g., \citealt{Mcginnis2018,Banzatti2019}). 

Based on these considerations, for sources with low inclination angles and small $R_K(OI)$ values (e.g., $< 0.5$) the \oi\ line broadening might be dominated by gas acceleration in the wind. We should therefore consider these values with caution.

\begin{figure*}[!t]
\includegraphics[trim= 480 300 40 60,width=1.0\columnwidth, angle=180]{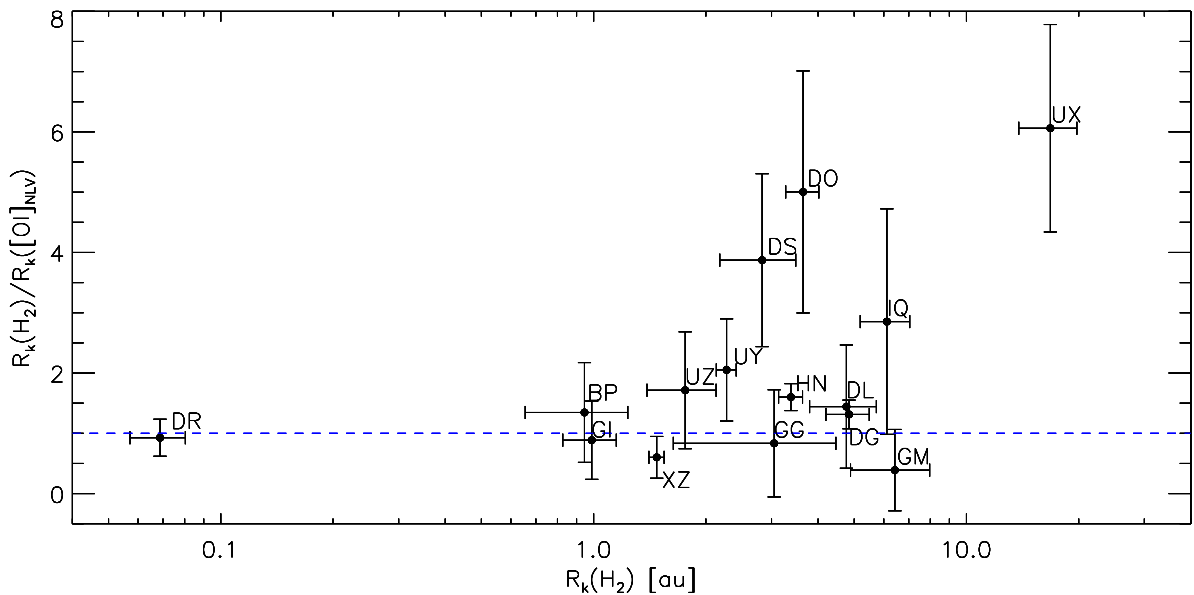}
\begin{center}\caption{\label{fig:radius_H2_ratio} $\rm R_K\text{(\htwo)}/R_{\rm k}(\text{\oi}_{\rm NLV})$ plotted against $\rm R_K\text{(\htwo)}$. $ R_{\rm k}$ is the disk radius where the \htwo\ or $\text{\oi}_{\rm NLV}$ component originates, under the assumption of Keplerian motion.}\end{center}
\end{figure*}

Noticeably,  in sources where $R_K({\rm OI})$ is estimated to be $\la$ 1 au,  the \htwo\  emission is never detected, except for DR Tau. The correlation between the emitting regions of \oi\ and \htwo\, is represented in Fig. \ref{fig:radius_H2_ratio}, where the $R_K({\rm H_2})/R_K({\rm OI})$ ratio is plotted against $R_K({\rm H_2})$ for sources where both lines have been detected.
This plot shows that the two emitting regions for most of the sources are equal to within a factor two. In only four objects (DS Tau, DO Tau, UX Tau, and IQ Tau) exceeds the emission region of \htwo\, that of \oi\, by factors between 3 and 6. Moreover, no appreciable differences in the ratio of emission sizes can be seen between sources that clearly have an extended wind and sources whose compact \htwo\ emission is consistent with an origin in a disk. 
We also remark that the transition disk sources with the largest inner cavities (GM Aur and UX Tau, see Table \ref{tab:H2properties_v2}) are also those with the largest estimated emitting region size. 

These results show that under the hypothesis of purely Keplerian broadening, the two emission regions are spatially connected in most of the cases. This might also explain the nondetection of \htwo\ when \oi\ is located near to the source: within a certain inner region, \htwo\ is more easily dissociated, and therefore the low column density of molecular gas prevents the detection of the 2.12 \um\ line (see also the discussion in the next section).
We also remark that the emission radius estimated for DR Tau (0.5 au for both \oi\ and \htwo) might be traced back to the values derived for the other sources in which \htwo\ has been detected if the inclination angle of the inner disk differs from the inclination of the outer disk measured with ALMA observations, as already discussed in Section 4.4.
Alternatively, the estimated radius may not be representative of the size of the \oi\ emission region, as discussed above, and it can be much larger than the measured value if the assumption of Keplerian broadening does not apply.

\subsection{Origin of the atomic and molecular emission}

In the previous sections we discussed the possible link between the \oi\ NLVC and \htwo\  emission suggested by their similar kinematic behavior. The two components  trace slightly different physical conditions, however.
The \htwo\ 2.12 \um\ line is typically excited at T$\sim$ 2000 - 6000 K \citep[e.g.,][]{Lepp1983}, and it is not particularly sensitive to the gas density. The \htwo\ near-IR ro-vibrational transitions can be collisionally excited in shocks or excited by direct absorption of high-energy photons from the star or from the accretion spots (UV or X-ray stimulated emission). As the \htwo\ is easily thermalized, it is not always easy to distinguish shocks from fluorescence emission from the ratio of different near-IR lines in relatively dense environments. Clear correlations between the \htwo\ 2.12 \um\, luminosity and X-ray emission have not been found (\citealt{Beck2019}), which indicates that excitation by high-energy photons is not a dominant mechanism in T Tauri disks. However, excitation by low-energy UV photons can still be compatible with observations (\citealt{Bary2003}). For DG Tau,  where the \htwo\ emission has been resolved and its origin identified in a wide-angle slow wind,  the molecular excitation has been attributed to either photoevaporation from the disk atmosphere or to ambipolar diffusion heating in a molecular MHD disk wind (\citealt{Agra-Amboage2014}), while shocks were found a less compelling mechanism. 

The excitation conditions of the \oi\ NLVC, on the other hand,  are still poorly constrained. Gas physical conditions have been generally studied on the whole LV component through line ratios of different optical/infrared forbidden lines (i.e., \citealt{Natta2014}, \citealt{Giannini2019}). These studies suggest gas temperatures in the range 5000-10000 K, high total densities (10$^6$-10$^7$ \cmt), and very low ionization fractions (x$_e << 0.1$). This last evidence indicates an origin in an almost neutral medium, at variance with the \oi\ HVC, where the ionization can reach high values depending on the jet velocity (\citealt{Giannini2019}). 
\cite{Fang2018} assessed the physical conditions in the LVC by separating the broad and narrow components, but found no appreciable difference between the two.
Based on these considerations, we can argue that there could be a continuity between the hot and neutral region of \oi\  emission and the warm molecular region where \htwo\ originates, and that 
the two species might overlap at the interface of these regions. 
Models of MHD disk winds might in principle be able to reproduce the above scenario: in these models, we would expect a decrease in the excitation conditions with increasing wind radius, in an onion-like structure composed of a faster or denser and highly excited material surrounded by a slower flow at low excitation. This scenario, if applied to \oi\ NLVC and \htwo\, , would imply that the variation in excitation conditions as a function of disk radius occurs on small scales, as we find very similar velocities between the two species and also a very similar launching radius in most of the cases. Noticeably, disk wind models that also address the thermochemical structure of the flow (\citealt{Panoglou2012}) predict that the coupling between charged and neutral fluids is sufficient to eject molecules from the disk out to at least 9 au. In these models, moreover, the launch radius beyond which most of the \htwo\ survives moves outward with evolutionary stage, and it is expected to be $>$ 1 au in class II T Tauri stars, which is in line with what we observe in our data.
Models for photoevaporative winds still do not consider the thermochemistry of the wind, and thus no prediction is so far available for \htwo\ emission in such winds. 

We note for the thermal structure of the photoevaporative models presented in Weber et al. (2020) that the temperatures at which the \htwo\ might be excited (i.e., 4000-6000 K) are predicted in a wide volume of the wind region in the radial and vertical direction. If \htwo\ originates from this region, then its emitting size would be larger than the size that is simply estimated from our assumption of Keplerian broadening. Although this cannot be disregarded in principle, it remains to be shown that these models are able to reproduce the observed similarities in FWHM and V$_p$ between \oi\ and \htwo\ through more complete chemical models that take exitation or dissociation of \htwo\ into account.

In a thorough analysis of the \htwo\ 2.12 \um\ wind emission in DG Tau, \cite{Agra-Amboage2014} addressed the hypothesis that the \htwo\ emission originates from a photoevaporative flow caused by the irradiation of the disk atmosphere. However, extreme conditions of irradiating far-UV flux, compatible with the DG Tau stellar flux, but not necessarily for  weaker sources, mean the case of DG Tau cannot easily be extrapolated to the other sources of our sample. 

Photoevaporative winds have also been suggested for the emission of the \oi\ NLVC. Models by \cite{Ercolano2016}  for a 0.7 M$_\odot$ star predict slightly blueshifted lines (2-5 \kms) with widths between 10 and 30 \kms, which are in line with what we measure in the majority of the sources. However, as discussed in Section 5.2, for sources with the highest blueshifted peak, the association with a prominent HVC would better favor the hypothesis that all the \oi\ components originate from the same MHD wind.

In sources where the \oi\ and \htwo\ emissions appear to originate from the same region ($R_K({\rm H_2})/R_K({\rm OI})\sim$ 1) and do not show an appreciable velocity shift (e.g., GM Aur and DR Tau),  this emission might in principle be also compatible with an origin from gas bound in the disk. Models of the thermochemical structure of the gaseous atmospheres of the inner disks in T Tauri stars show a stratification in depth of the temperature and gas dissociation. These models predict that neutral oxygen is abundant in the upper disk atmospheres, but that a transitional zone between the upper hot layer and the cold disk mid-plane should be present at temperatures up to 2000 K, where \htwo\ ro-vibrational transitions might be excited (\citealt{Glassgold2004}). We note, however, that the line fluxes measured in our sources are at least a factor of 10 higher than those predicted by models of \htwo\, excitation in protoplanetary disks by X-ray and UV-irradiation (\citealt{Nomura2007}) when their predictions are scaled to the Taurus distance. 

We remark that  \htwo\ fluorescent emission in the UV wavelength range has been observed by the Hubble Space Telescope in some of the stars of our sample. Most of the UV \htwo\ emission appears to originate in disks (\citealt{France2012}). Modeling of the line profiles indicates that the radial distribution of this emission extends from $\la$ 0.1 au to $\sim$ 10 au, that is, down to inner regions at smaller radii than the near-IR emission discussed here (Hoadley et al. 2015).
This could be explained by a higher efficiency of the  Ly$\alpha$-pumping mechanism to excite the fluorescent lines when \htwo\ is closer to the UV photons, coupled with the higher sensitivity of the UV observations made with the Hubble Space Telescope with respect to near-IR observations.

Finally, we also note that molecular emission due to CO and H$_2$O in the near-IR (2-5 \um) is also observed in T Tauri stars, and the emission regions implied by their resolved profiles are in the range 0.03-10 au. This means that these molecules survive closer to the central star than the warm \htwo\ giving rise to the 2.12 \um\ emission. 
This is consistent with emission of these molecules in the disk: they originate from a deeper and colder (T about 1000 K) disk layer, where 
dissociation from high-energy photons is prevented and their abundance can be maintained at a relatively high level even at very close distance to the central star. In these regions, only the \htwo\ mid-IR pure rotational lines would be excited,  while the temperature is not sufficiently high to significantly excite the 2.12 \um\ line (\citealt{Glassgold2009}). 

In conclusion, our findings support a scenario where the \oi\ NLVC and \htwo\ both originate mostly in low-velocity winds, and that the near-IR \htwo\ emission in these winds is only excited to a detectable level when this wind is driven from disk regions larger than 1 au. With the observations we presented, it is not possible without proper predictions of line profiles by models that simultaneously address the physical and chemical structure to distinguish photoevaporative from MHD disk wind models.

Emission in disks seems to be a less compelling reason because of the brightness of the line emission with respect to expectation from existing models. However, this possibility cannot be ruled out for sources with v$\rm_p \sim 0$ and whose \oi\ emission is not dominated by HV gas. This needs to be 
better explored by means of tailored model comparisons.

\section{Conclusions}

We have presented the analysis of the \oi\ 630 nm and \htwo\ 2.12 \um\ lines observed at resolutions of $\sim$ 3 \kms\ and 6 \kms\, , respectively, in a sample of 36 classical T Tauri stars of the Taurus-Auriga star-forming region. The spectra were flux calibrated using photometric observations acquired as close as possible in time. \htwo\ 2.12 \um\, is detected in 17 sources ($\sim$ 50\% of detection) and \oi\ is detected in all sources but one.
We applied a Gaussian decomposition of the line profiles to separate different kinematic components. The \htwo\ line profile was fit for most of the sources as a single Gaussian, although for about one-third of the targets, a second weaker and more strongly blueshifted component was also identified. The \oi\ line profile can be deconvolved into components at different velocities, in line with what has been found in previous high-resolution studies.  We concentrated in particular on the most frequently detected narrow component at the lowest velocity (v$\rm_p < 20$ \kms), that is, the narrow low-velocity component (NLVC). The main results of our study are summarized below.

We found a strong kinematic link between the \htwo\ 2.12 \um\ and the NLVC of the \oi\ emission. In particular, the peak velocities and the FWHM of the two lines are tightly correlated, suggesting that molecular and neutral atomic emission originate from closely related regions.

In seven sources, \htwo\ and \oi\, have clearly blueshifted peaks that indicate an origin in winds. These sources all have prominent \oi\ high-velocity emission or known resolved jets. On the other hand, a strong \oi\ HVC is also observed in many sources with no appreciable centroid shift, which leads to the conclusion that jets do not necessarily affect the kinematics of the low-velocity winds. In only two sources (GM Aur and DS Tau) do \oi\ and \htwo\ have a single unshifted component: this emission is therefore consistent with an origin in the disk, as also suggested for GM Aur from previous high-angular resolution observations (\citealt{Beck2019}). We note, however, that GM Aur presents a transition disk, and the lack of a blueshift could also be explained by emission from the receding part that reaches the observer through the cavity of the disk, as modeled by \citet{Ercolano2010}.

We do not observe any clear correlation between v$\rm_p$ of the \htwo\, and \oi\, lines and the disk inclination, in line with previous studies. However, the low velocity exhibited by most of the sources, close to our resolution limit, makes the observation of any correlation challenging.

Assuming that the line width is dominated by Keplerian broadening, we measured the radius of the region in the disk where the emission originates. We find that the \oi\ NLVC originates from 0.05 and 20 au and \htwo\, from 2 and 20 au. The emission regions of \htwo\, and \oi\, are comparable in size for sources where both the lines are detected. Noticeably, \htwo\ is never detected in sources where $R_K(OI)$ is below 1 au. 

This finding, together with the additional evidence that \htwo\ 2.12 \um\ is never detected in sources of earlier spectral type ($\sim$ F-G) with a luminosity $>1$ \lsun, suggests that the survival of warm \htwo\ in disks strongly depends on the gas exposure to the radiation from the central star. 

These pieces of evidence suggest that the \oi\ NLVC and the \htwo\ emitting regions might overlap although different conditions hold for the two species (temperature and dissociation). The most likely scenario is that the two species are part of the same wind, where radial gradients of excitation conditions occur on small spatial scales, without appreciable velocity variations. If this is the case, our observations suggest that winds originating from luminous stars or from the very inner regions do not have a strong warm molecular counterparts. From observations alone we cannot distinguish whether the wind giving rise to both emissions is thermally excited (photoevaporation) or due to the action of a magnetic field (MHD wind). Models of line excitation and profiles including molecular chemistry are needed to distinguish the two scenarios. However, we remark that thermochemical models of MHD disk winds predict that in class II sources, the launch radii beyond which \htwo\ survives is $>1$ au (\citealt{Panoglou2012}), in line with our findings. 

For the few sources of our sample in which the \oi\ emission is dominated by a single NLVC (e.g., GM Aur and DS Tau), an origin from gas bound in the disk cannot be ruled out. In this case, \oi\ and \htwo\ might be lifted from the disk at different depths, that is, \oi\, could come from the upper disk atmosphere, where the gas temperature is about 5000 K, while \htwo\ could originate from a deeper transition zone with a temperature of up to 2000 K, whose thickness and location depend on the surface heating (\citealt{Glassgold2004}).

Our results highlight the importance of high-resolution spectroscopy over a wide wavelength range in order to link tracers of different manifestations of the same phenomenon. Future directions of our study will be the possibility to use sensitive IR high-resolution spectrometers, such as CRIRES on the Very Large Telescope, to explore the \htwo\ emission of sources below our present detection limit, and simultaneously compare the warm molecular gas traced by the \htwo\ 2.12 \um\ with other molecular tracers (e.g., CO and H$_2$O). At the same time, we note the high demand for suitable models that can correctly interpret the large amount of information that is being gathered on physical and chemical properties of winds in protoplanetary disks.

\begin{acknowledgements}
We thanks Dr. Andrea Di Paola for his assistance during the observations at Campo Imperatore. We acknowledge the 'Asiago Novae and Symbiotic Stars' Collaboration for their support in the optical low-resolution observations. This work  has  been supported  by the project PRIN-INAF-MAIN-STREAM 2017 “Protoplanetary disks seen through the eyes of new-generation instruments”. We also  acknowledge the support by INAF/Frontiera through the ‘Progetti Premiali’ funding scheme of the Italian Ministry of Education, University, and Research, by the European Union's Horizon 2020 research and innovation program under the Marie Sklodowska-Curie grant agreement No 823823 (DUSTBUSTERS) and by the Deutsche Forschungs-Gemeinschaft (DFG, German Research Foundation) - Ref no. FOR 26341/1 TE 1024/1.1
\end{acknowledgements}

%
%

\begin{appendix}
\section{}
\begin{figure*}[!t]
\includegraphics[trim= 600 440 60 60,width=0.52\columnwidth, angle=180]{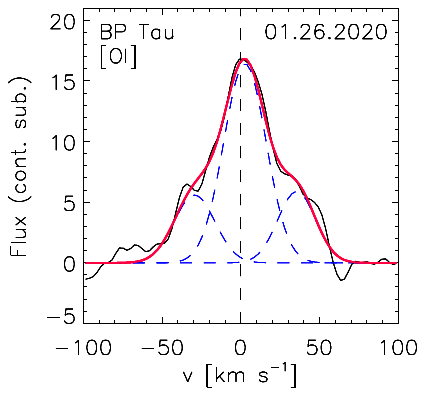}
\includegraphics[trim= 600 440 60 60,width=0.52\columnwidth, angle=180]{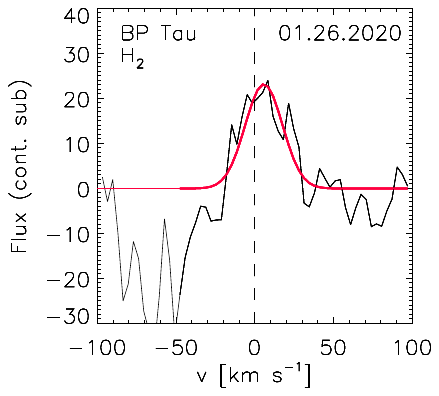}
\includegraphics[trim= 600 440 60 60,width=0.52\columnwidth, angle=180]{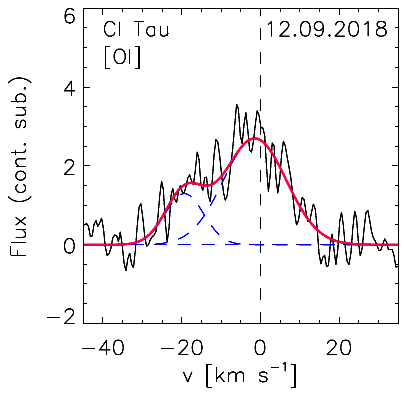}
\includegraphics[trim= 600 440 60 60,width=0.52\columnwidth, angle=180]{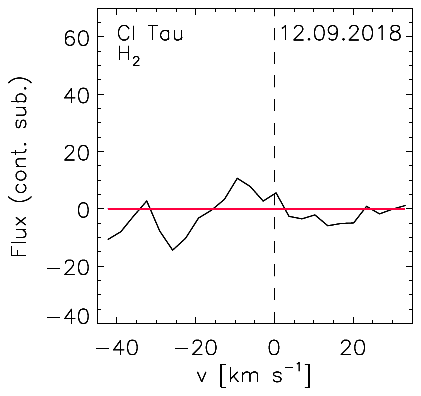}
\includegraphics[trim= 600 440 60 60,width=0.52\columnwidth, angle=180]{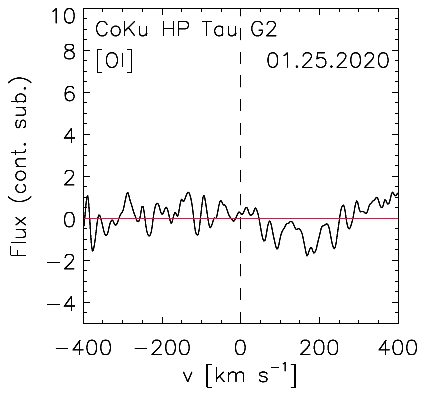}
\includegraphics[trim= 600 440 60 60,width=0.52\columnwidth, angle=180]{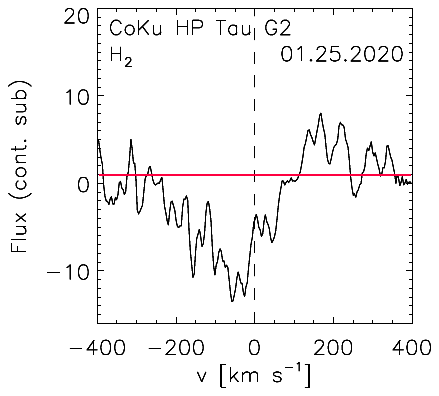}
\includegraphics[trim= 600 440 60 60,width=0.52\columnwidth, angle=180]{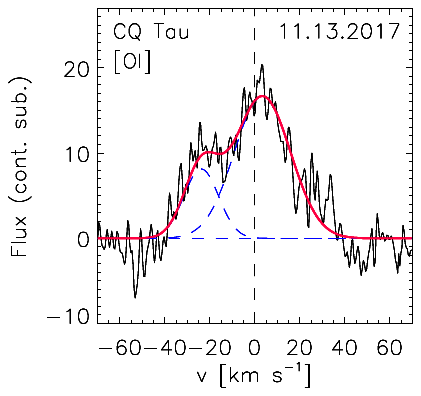}
\includegraphics[trim= 600 440 60 60,width=0.52\columnwidth, angle=180]{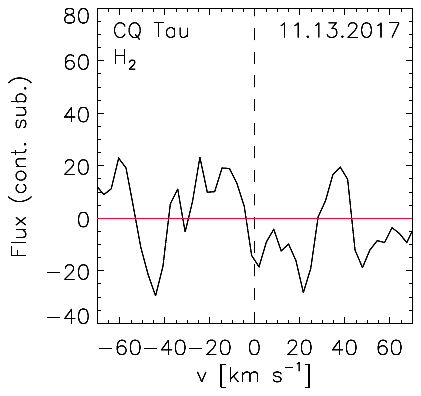}
\includegraphics[trim= 600 440 60 60,width=0.52\columnwidth, angle=180]{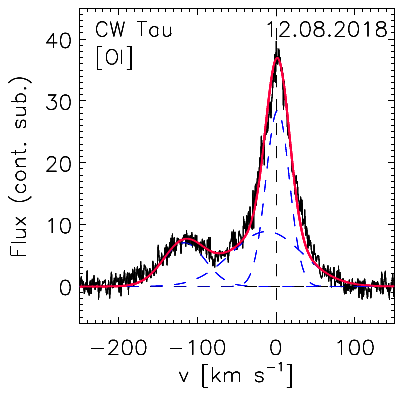}
\includegraphics[trim= 600 440 60 60,width=0.52\columnwidth, angle=180]{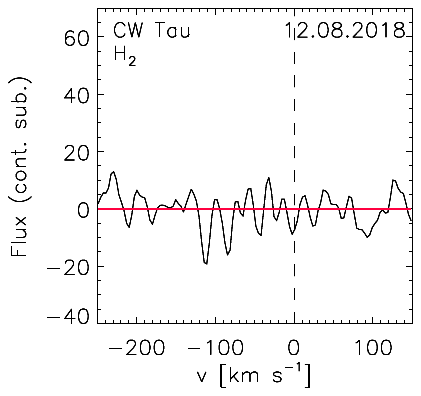}
\includegraphics[trim= 600 440 60 60,width=0.52\columnwidth, angle=180]{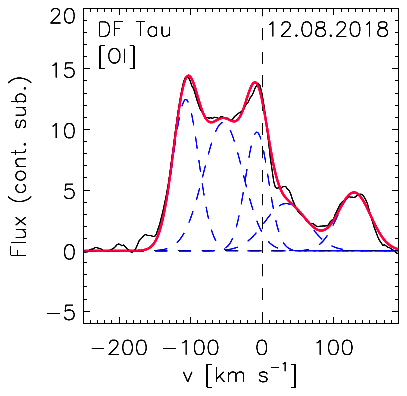}
\includegraphics[trim= 600 440 60 60,width=0.52\columnwidth, angle=180]{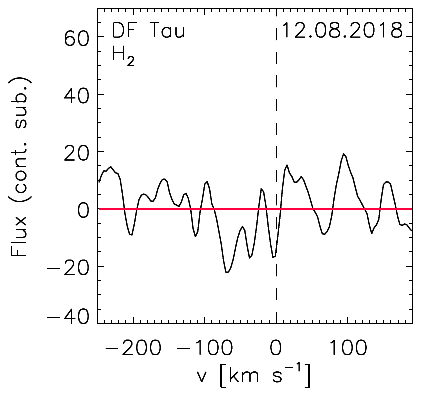}
\includegraphics[trim= 600 440 60 60,width=0.52\columnwidth, angle=180]{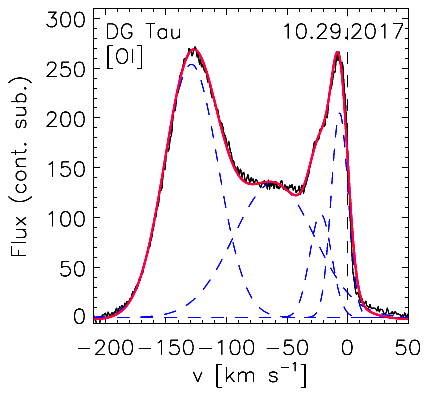}
\includegraphics[trim= 600 440 60 60,width=0.52\columnwidth, angle=180]{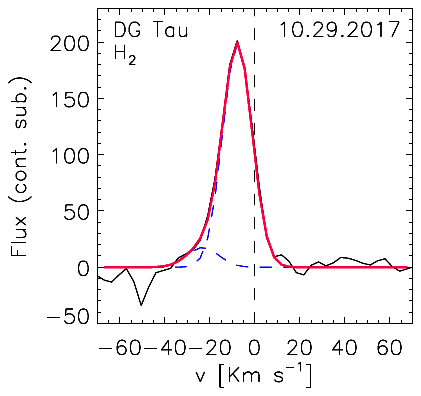}
\includegraphics[trim= 600 440 60 60,width=0.52\columnwidth, angle=180]{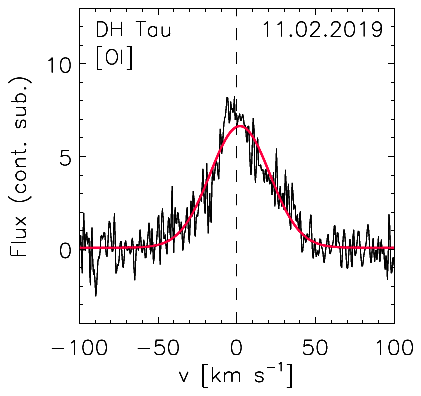}
\includegraphics[trim= 600 440 60 60,width=0.52\columnwidth, angle=180]{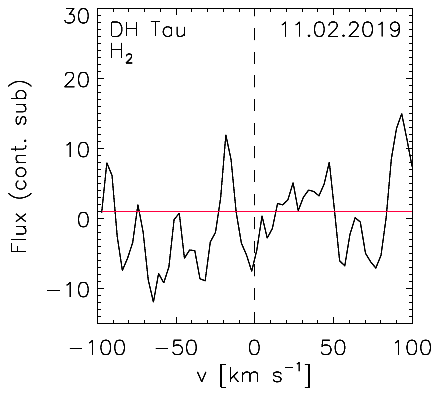}
\includegraphics[trim= 600 440 60 60,width=0.52\columnwidth, angle=180]{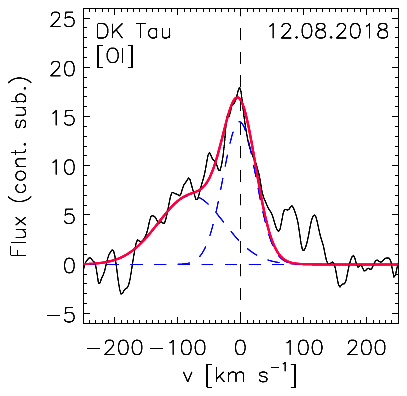}
\includegraphics[trim= 600 440 60 60,width=0.52\columnwidth, angle=180]{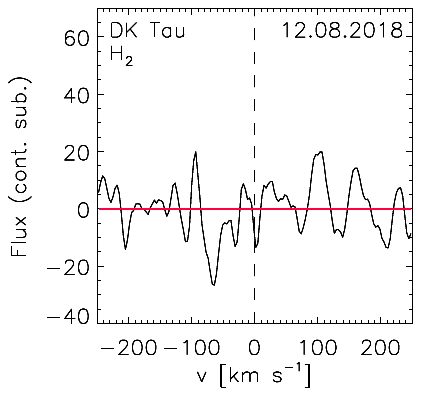}
\includegraphics[trim= 600 440 60 60,width=0.52\columnwidth, angle=180]{Figures/OI_profile/DL_Tau_OI.pdf}
\includegraphics[trim= 600 440 60 60,width=0.52\columnwidth, angle=180]{Figures/H2_profile/DL_Tau_H2.pdf}
\includegraphics[trim= 600 440 60 60,width=0.52\columnwidth, angle=180]{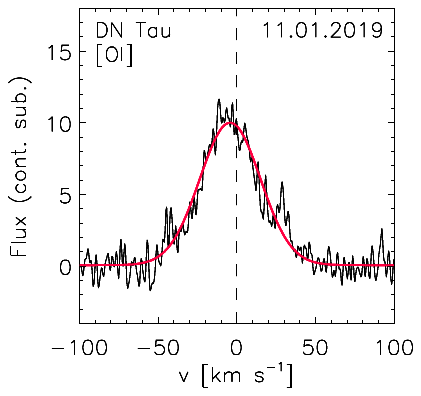}
\includegraphics[trim= 600 440 60 60,width=0.52\columnwidth, angle=180]{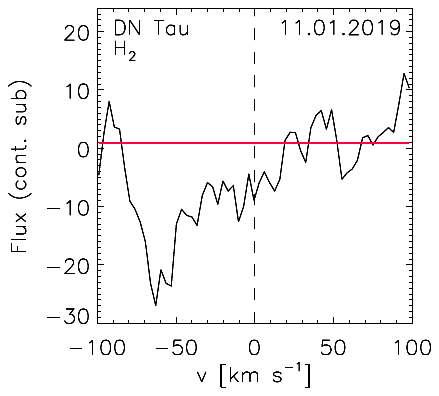}
\includegraphics[trim= 600 440 60 60,width=0.52\columnwidth, angle=180]{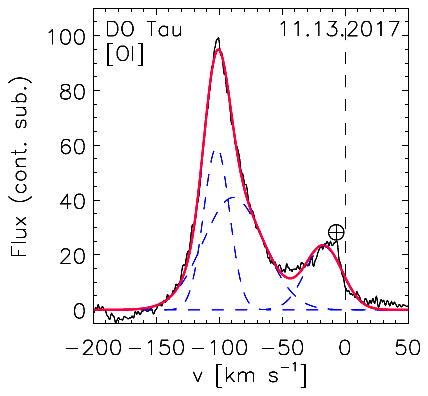}
\includegraphics[trim= 600 440 60 60,width=0.52\columnwidth, angle=180]{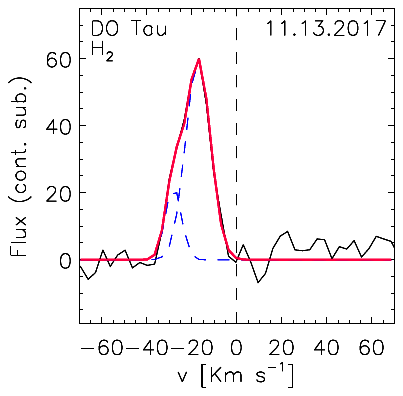}
\begin{center}\caption{\label{fig:OI_H2_profile} Continuum-subtracted \oi\ and \htwo\ line profiles. In red we plot the fit to the profile, obtained by adding single or multiple Gaussians (dashed blue lines). Flux units are $10^{-15}$ $\rm erg s^{-1}$ $\rm cm^{-2}$ $\AA^{-1}$. For each panel, we indicate the target name and date of observation (MM.DD.YYYY).}\end{center}
\end{figure*}

\begin{figure*}[!t]
\includegraphics[trim= 600 440 60 60,width=0.52\columnwidth, angle=180]{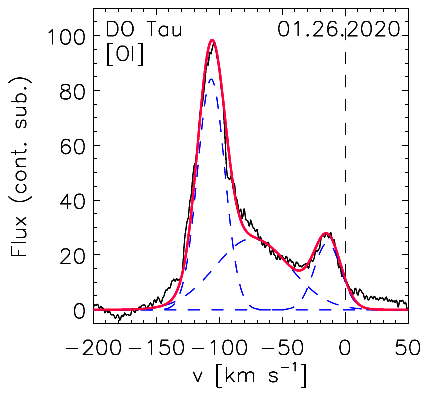}
\includegraphics[trim= 600 440 60 60,width=0.52\columnwidth, angle=180]{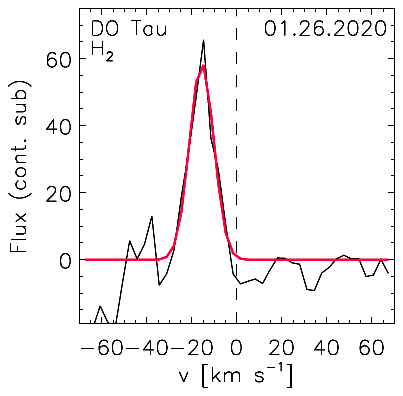}
\includegraphics[trim= 600 440 60 60,width=0.52\columnwidth, angle=180]{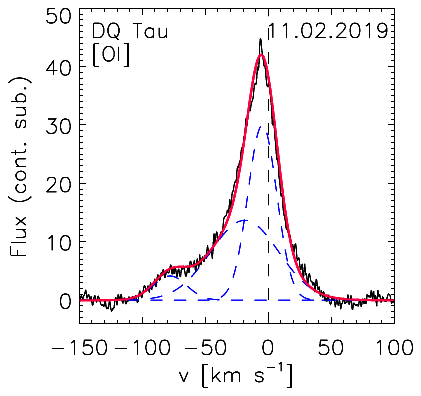}
\includegraphics[trim= 600 440 60 60,width=0.52\columnwidth, angle=180]{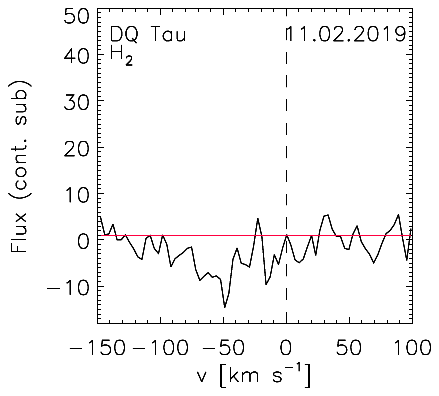}
\includegraphics[trim= 600 440 60 60,width=0.52\columnwidth, angle=180]{Figures/OI_profile/DR_Tau_OI.pdf}
\includegraphics[trim= 600 440 60 60,width=0.52\columnwidth, angle=180]{Figures/H2_profile/DR_Tau_H2.pdf}
\includegraphics[trim= 600 440 60 60,width=0.52\columnwidth, angle=180]{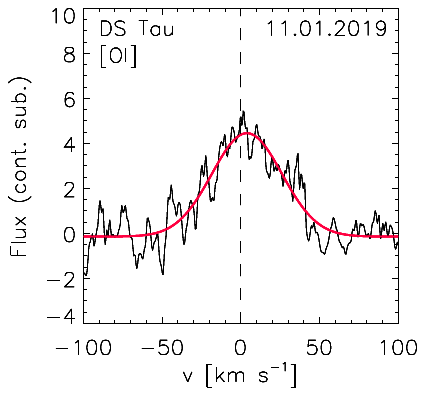}
\includegraphics[trim= 600 440 60 60,width=0.52\columnwidth, angle=180]{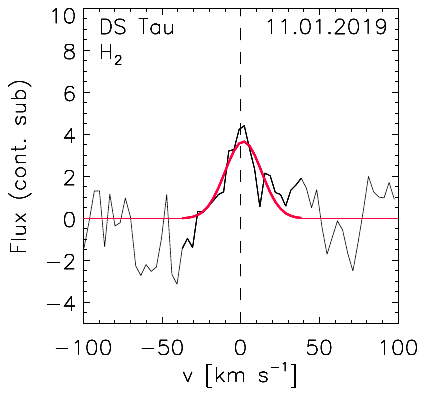}
\includegraphics[trim= 600 440 60 60,width=0.52\columnwidth, angle=180]{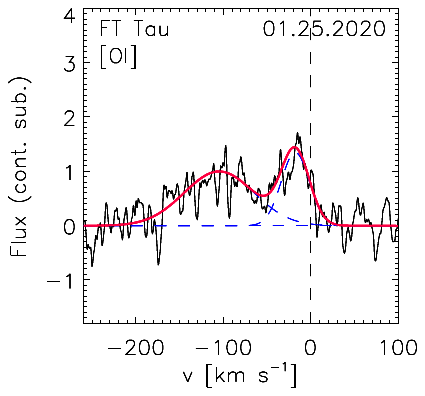}
\includegraphics[trim= 600 440 60 60,width=0.52\columnwidth, angle=180]{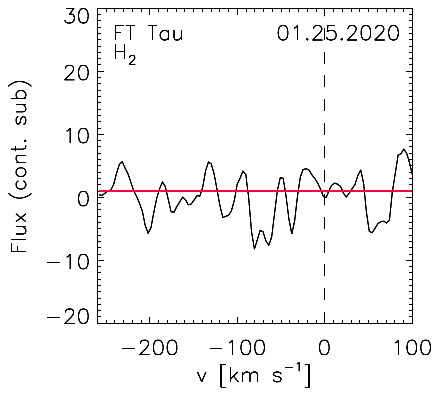}
\includegraphics[trim= 600 440 60 60,width=0.52\columnwidth, angle=180]{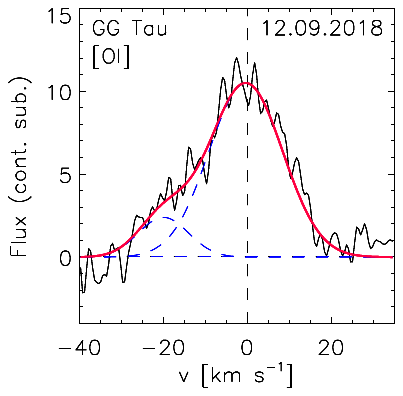}
\includegraphics[trim= 600 440 60 60,width=0.52\columnwidth, angle=180]{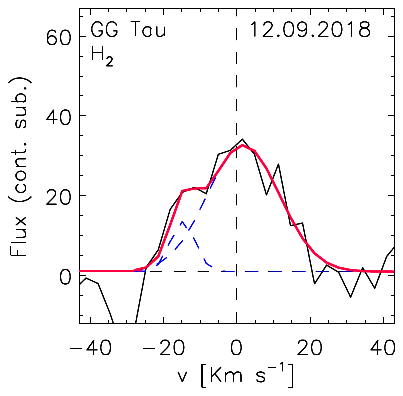}
\includegraphics[trim= 600 440 60 60,width=0.52\columnwidth, angle=180]{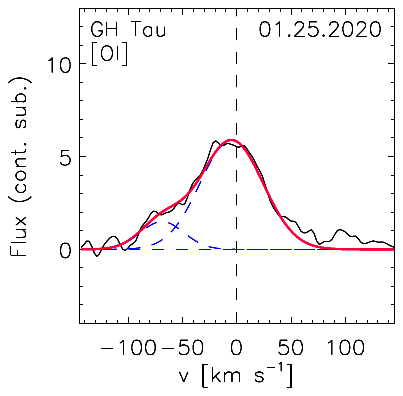}
\includegraphics[trim= 600 440 60 60,width=0.52\columnwidth, angle=180]{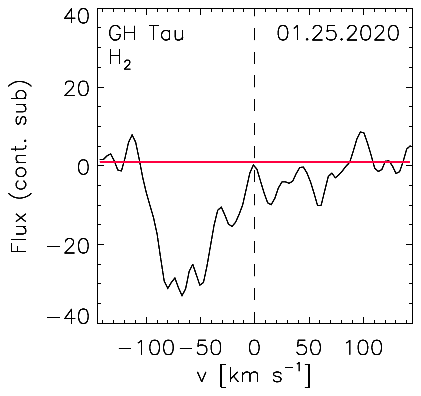}
\includegraphics[trim= 600 440 60 60,width=0.52\columnwidth, angle=180]{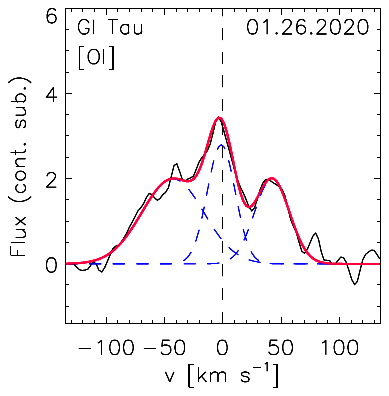}
\includegraphics[trim= 600 440 60 60,width=0.52\columnwidth, angle=180]{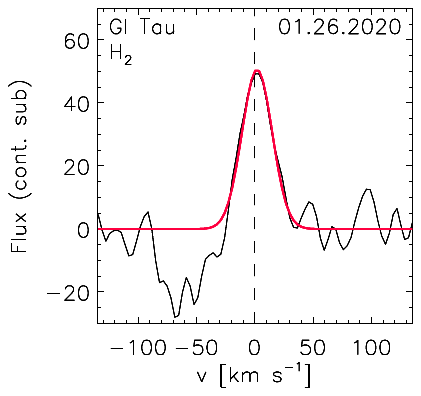}
\includegraphics[trim= 600 440 60 60,width=0.52\columnwidth, angle=180]{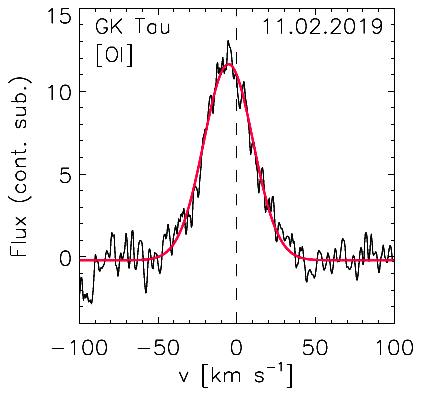}
\includegraphics[trim= 600 440 60 60,width=0.52\columnwidth, angle=180]{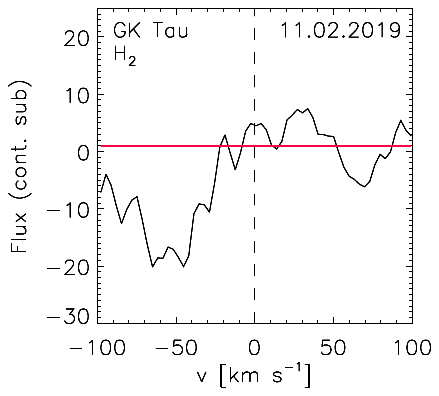}
\includegraphics[trim= 600 440 60 60,width=0.52\columnwidth, angle=180]{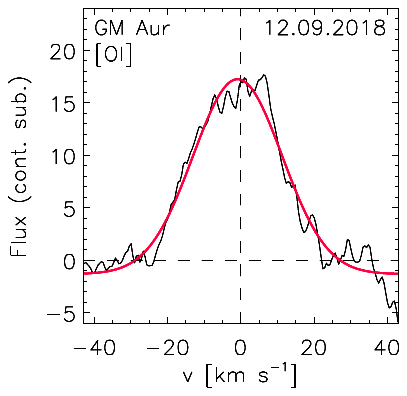}
\includegraphics[trim= 600 440 60 60,width=0.52\columnwidth, angle=180]{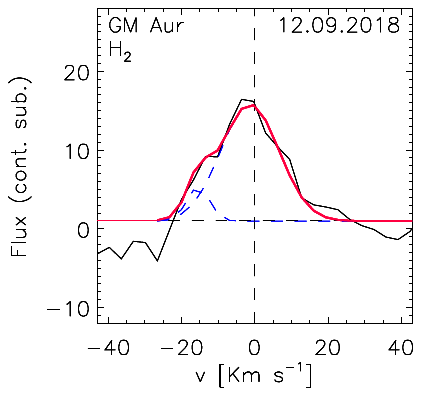}
\includegraphics[trim= 600 440 60 60,width=0.52\columnwidth, angle=180]{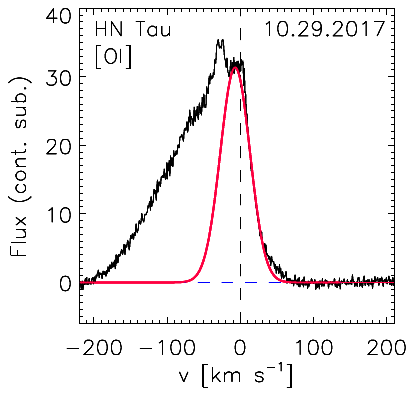}
\includegraphics[trim= 600 440 60 60,width=0.52\columnwidth, angle=180]{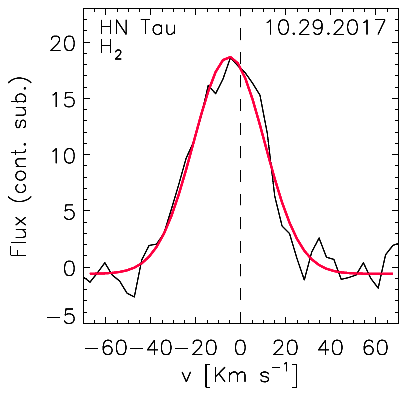}
\includegraphics[trim= 600 440 60 60,width=0.52\columnwidth, angle=180]{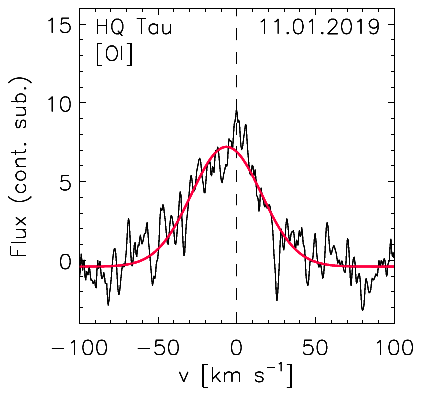}
\includegraphics[trim= 600 440 60 60,width=0.52\columnwidth, angle=180]{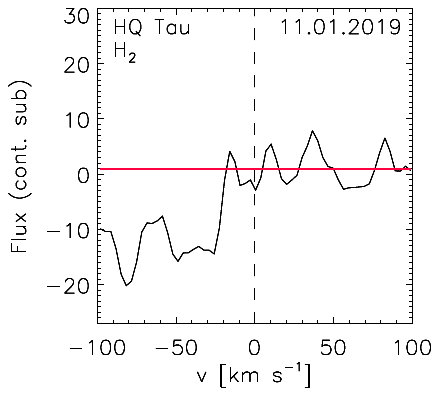}

\begin{center} \textbf{Fig. 1.} (Continued).\end{center}
\end{figure*}

\begin{figure*}[!t]
\includegraphics[trim= 600 440 60 60,width=0.52\columnwidth, angle=180]{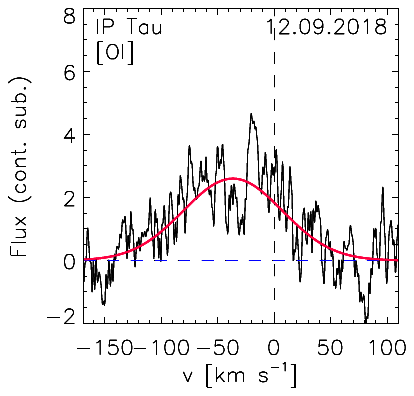}
\includegraphics[trim= 600 440 60 60,width=0.52\columnwidth, angle=180]{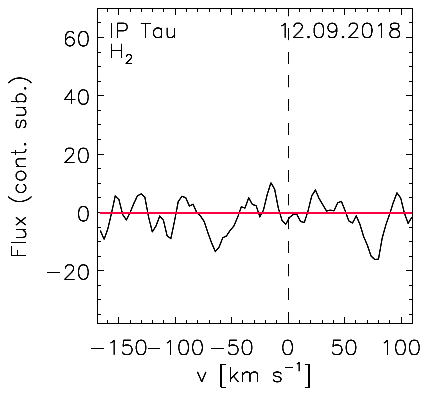}
\includegraphics[trim= 600 440 60 60,width=0.52\columnwidth, angle=180]{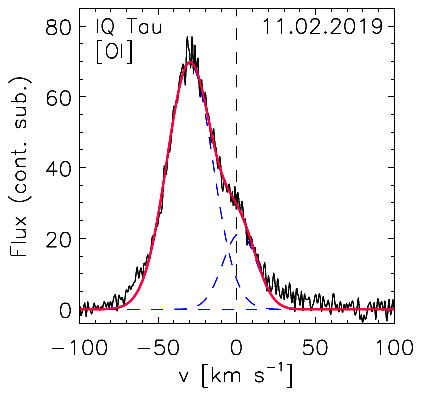}
\includegraphics[trim= 600 440 60 60,width=0.52\columnwidth, angle=180]{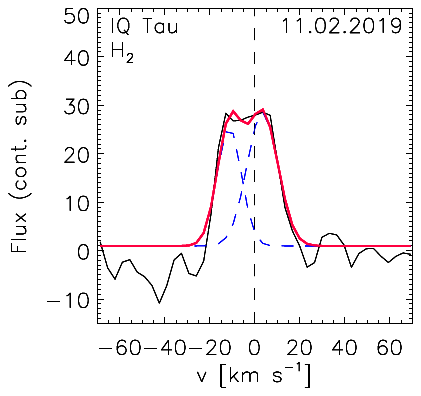}
\includegraphics[trim= 600 440 60 60,width=0.52\columnwidth, angle=180]{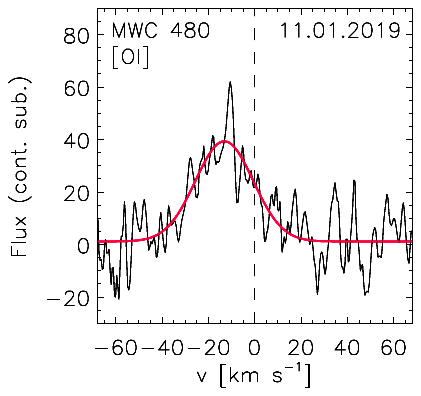}
\includegraphics[trim= 600 440 60 60,width=0.52\columnwidth, angle=180]{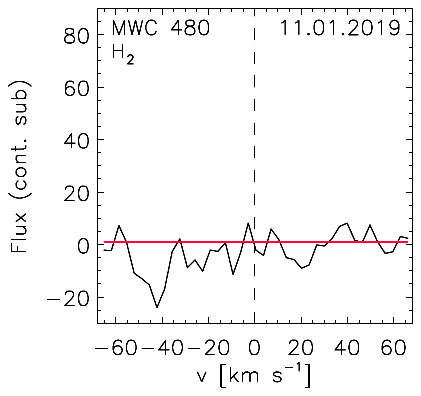}
\includegraphics[trim= 600 440 60 60,width=0.52\columnwidth, angle=180]{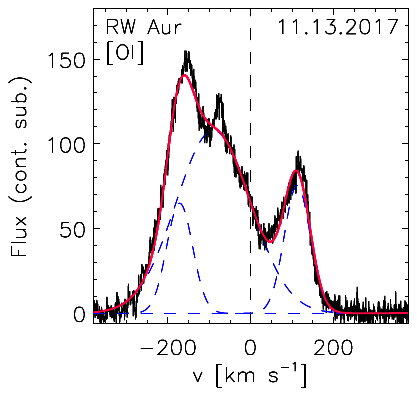}
\includegraphics[trim= 600 440 60 60,width=0.52\columnwidth, angle=180]{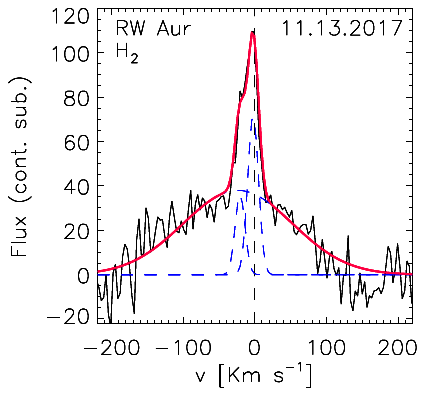}
\includegraphics[trim= 600 440 60 60,width=0.52\columnwidth, angle=180]{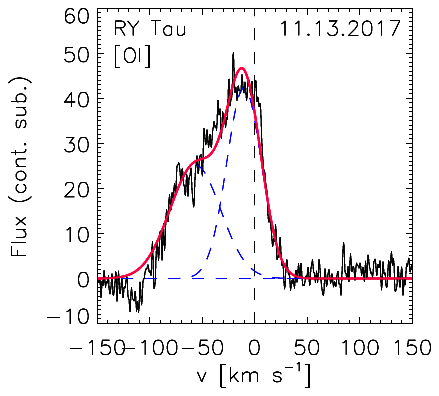}
\includegraphics[trim= 600 440 60 60,width=0.52\columnwidth, angle=180]{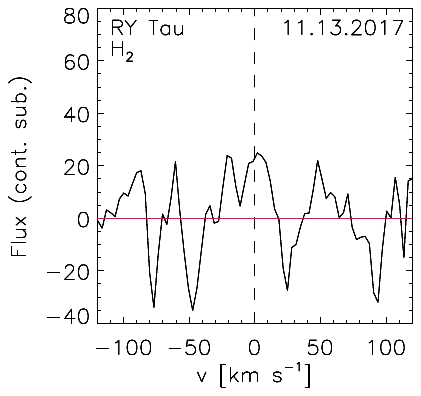}
\includegraphics[trim= 600 440 60 60,width=0.52\columnwidth, angle=180]{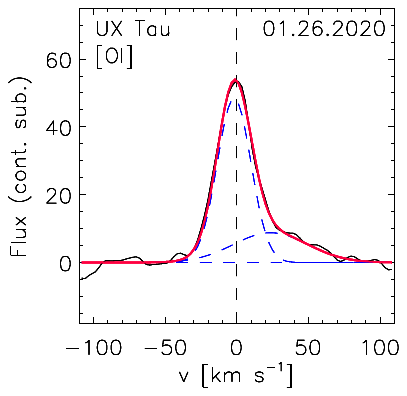}
\includegraphics[trim= 600 440 60 60,width=0.52\columnwidth, angle=180]{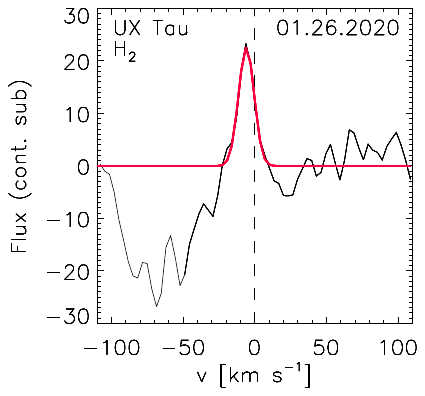}
\includegraphics[trim= 600 440 60 60,width=0.52\columnwidth, angle=180]{Figures/OI_profile/UY_Aur_OI.pdf}
\includegraphics[trim= 600 440 60 60,width=0.52\columnwidth, angle=180]{Figures/H2_profile/UY_Aur_H2.pdf}
\includegraphics[trim= 600 440 60 60,width=0.52\columnwidth, angle=180]{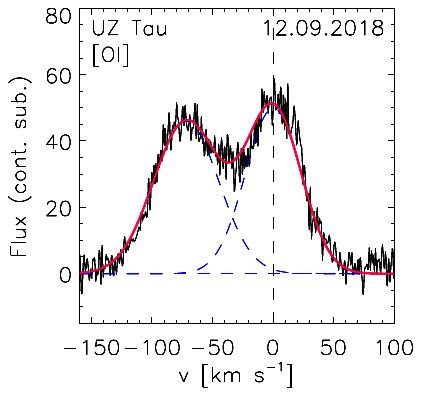}
\includegraphics[trim= 600 440 60 60,width=0.52\columnwidth, angle=180]{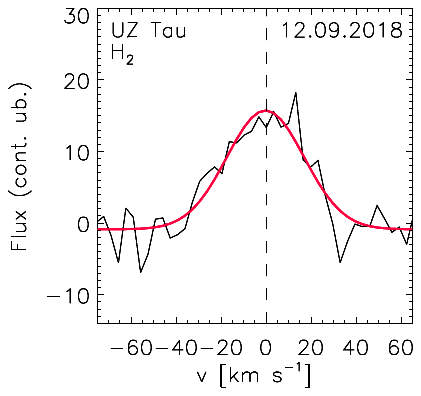}
\includegraphics[trim= 600 440 60 60,width=0.52\columnwidth, angle=180]{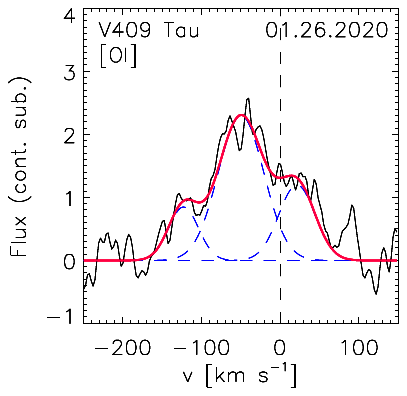}
\includegraphics[trim= 600 440 60 60,width=0.52\columnwidth, angle=180]{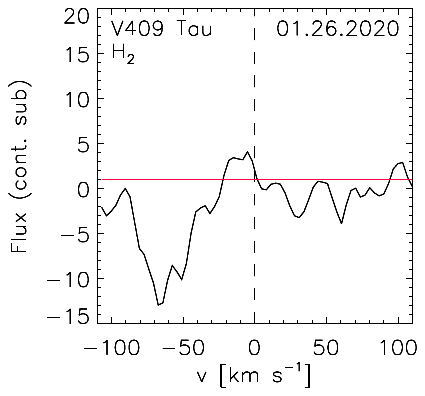}
\includegraphics[trim= 600 440 60 60,width=0.52\columnwidth, angle=180]{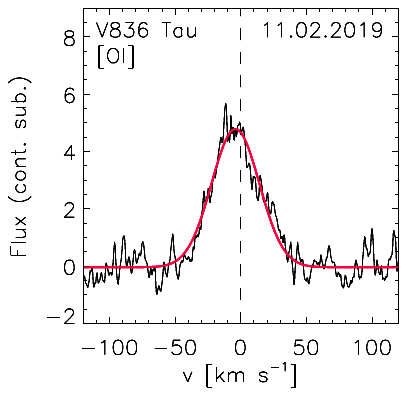}
\includegraphics[trim= 600 440 60 60,width=0.52\columnwidth, angle=180]{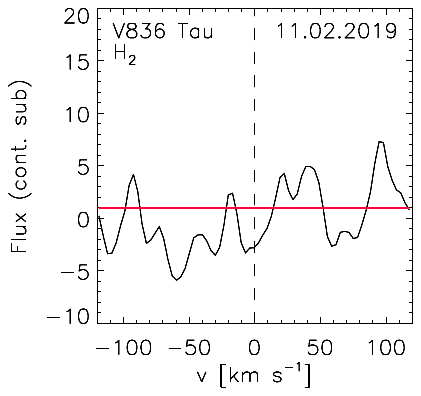}
\includegraphics[trim= 600 440 60 60,width=0.52\columnwidth, angle=180]{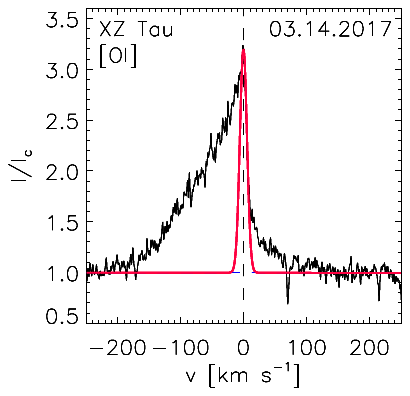}
\includegraphics[trim= 600 440 60 60,width=0.52\columnwidth, angle=180]{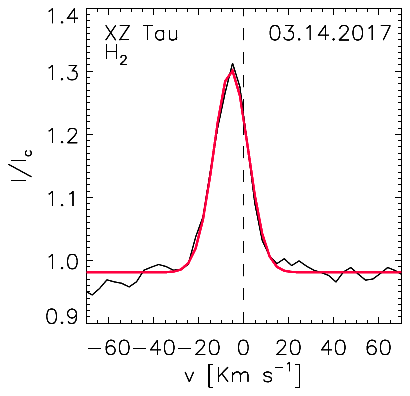}
\includegraphics[trim= 600 440 60 60,width=0.52\columnwidth, angle=180]{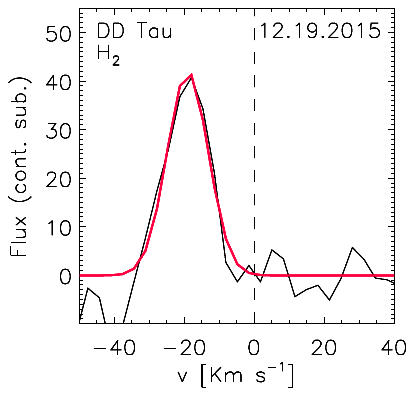}
\includegraphics[trim= 600 440 60 60,width=0.52\columnwidth, angle=180]{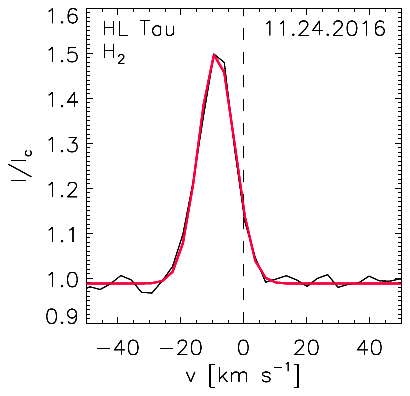}
\begin{center} \textbf{Fig. 1.} (Continued).\end{center}
\end{figure*}

\begin{table*}
\small
\center
\caption{\label{tab:KinematicParametersAll} Line kinematic parameters of the Gaussian components having $\rm |v|>30$ $\rm km$ $\rm s^{-1}$ for the \oi\ and \htwo\ profiles.}
\begin{tabular}{lc|ccc|ccc}
\hline
\hline
          &             & \multicolumn{3}{c}{\oi}                     & \multicolumn{3}{c}{\htwo}		                                 \\	
Source    & Obs Date    & $\rm v_p$               & $\rm FWHM$              & $\rm I_p$ ($\times 10^{-15}$)    & $\rm v_p$     & $\rm FWHM$       & $\rm I_p$ ($\times 10^{-15}$)          \\
          &             & [$\rm km$ $\rm s^{-1}$] & [$\rm km$ $\rm s^{-1}$] & [$\rm erg$ $\rm s^{-1}$ $\rm cm^{-2}$ $\AA^{-1}$] & [$\rm km$ $\rm s^{-1}$] & [$\rm km$ $\rm s^{-1}$] & [$\rm erg$ $\rm s^{-1}$ $\rm cm^{-2}$ $\AA^{-1}$]                    \\    
\hline   
 BP Tau   & 01-26-2020  & 35.4   $\pm$ 6.5  & 28.1  $\pm$ 12.9  & 5.9  $\pm$ 2.1  & ND                & ND               & ND                \\  
          &             & -29.5  $\pm$ 5.8  & 30.5  $\pm$ 19.5  & 5.6  $\pm$ 2.0  & ND                & ND               & ND                \\  
 CI Tau   & 12-09-2018  & -19.4  $\pm$ 1.8  & 11.2  $\pm$ 5.6   & 1.3  $\pm$ 0.3  & ND                & ND               & ND                \\
 CoKu HP  & 01-25-2020  & ND                & ND                & ND              & ND                & ND               & ND                \\  
 CQ Tau   & 11-13-2017  & -23.9  $\pm$ 3.9  & 18.1  $\pm$  6.7  & 81.5 $\pm$ 33.3 & ND                & ND               & ND                \\
 CW Tau   & 12-08-2018  & -12.2  $\pm$ 17.3 & 105.9 $\pm$ 39.5  & 8.9  $\pm$ 2.6  & ND                & ND               & ND                \\  
          &             & -117.0 $\pm$ 13.2 & 62.3  $\pm$ 45.8  & 7.1  $\pm$ 2.8  & ND                & ND               & ND                \\   
 DD Tau   & 12-19-2015  & -                 & -                 & -               & ND                & ND               & ND                \\
 DF Tau   & 12-08-2018  & 128.1  $\pm$ 5.2  & 54.1  $\pm$ 13.5  & 4.8  $\pm$ 1.0  & ND                & ND               & ND                \\
          &             & 32.9   $\pm$ 13.2 & 70.6  $\pm$ 20.3  & 3.9  $\pm$ 1.0  & ND                & ND               & ND                \\
          &             & -54.8  $\pm$ 3.0  & 64.7  $\pm$ 5.5   & 10.6 $\pm$ 1.0  & ND                & ND               & ND                \\
          &             & -107.1 $\pm$ 1.5  & 42.3  $\pm$ 6.7   & 12.5 $\pm$ 1.2  & ND                & ND               & ND                \\
 DG Tau   & 12-19-2015  & -                 & -                 & -               & -32.5 $\pm$ 5.3   & 11.9 $\pm$ 9.2   & 45.0 $\pm$ 26.0   \\
          & 10-29-2017  & -22.5  $\pm$ 0.6  & 21.0  $\pm$ 0.9   & 104.0 $\pm$ 3.5 & -23.2 $\pm$ 8.5   & 15.5 $\pm$ 13.2  & 17.5 $\pm$ 13.7   \\
          &             & -61.0  $\pm$ 0.8  & 77.7  $\pm$ 0.9   & 135.5 $\pm$ 2.2 & ND                & ND               & ND                \\
          &             & -129.3 $\pm$ 1.1  & 52.0  $\pm$ 0.6   & 254.0 $\pm$ 1.7 & ND                & ND               & ND                \\
 DH Tau   & 11-02-2019  & ND                & ND                & ND              & ND                & ND               & ND                \\
 DK Tau   & 12-08-2018  & -79.1 $\pm$ 20.1  & 122.4 $\pm$ 51.6  & 7.0 $\pm$ 2.4   & ND                & ND               & ND                \\    
 DL Tau   & 12-20-2015  & -                 & -                 & -                 & -12.7 $\pm$ 7.6 & 7.1 $\pm$ 7.5    & 21.5 $\pm$ 20.1   \\
          & 10-29-2017  & 20.0 $\pm$ 21.7   & 38.5  $\pm$ 36.6  & 1.5  $\pm$ 0.9  & -17.9   $\pm$ 5.2 & 10.3 $\pm$ 7.6   & 18.0 $\pm$ 11.6   \\
          &             & -18.8 $\pm$ 6.9   & 13.4  $\pm$ 21.0  & 3.1  $\pm$ 1.3  & ND                & ND               & ND                \\
          &             & -68.0 $\pm$ 13.7  & 94.1  $\pm$ 31.6  & 2.7  $\pm$ 0.6  & ND                & ND               & ND                \\
          &             & -164.5 $\pm$ 11.9 & 85.9  $\pm$ 19.2  & 3.4  $\pm$ 8.7  & ND                & ND               & ND                \\             
 DN Tau   & 11-01-2019  & ND                & ND                & ND              & ND                & ND               & ND                \\           
 DO Tau   & 12-19-2015  & -                 & -                 & -               & -13.4 $\pm$ 2.8   & 7.1 $\pm$ 2.5    & 19.5 $\pm$ 7.2    \\
          & 11-13-2017  & -89.0 $\pm$ 1.8   & 56.4  $\pm$ 2.4   & 41.0 $\pm$ 1.4  & -28.3 $\pm$ 4.6   & 5.7 $\pm$ 4.8    & 22.0 $\pm$ 18.0   \\
          &             & -102.3 $\pm$ 0.7  & 24.0  $\pm$ 1.1   & 59.3 $\pm$ 2.1  & ND                & ND               & ND                \\
          & 01-26-2020  & -74.0 $\pm$ 3.9   & 69.4  $\pm$ 14.2  & 26.0 $\pm$ 2.9  & ND                & ND               & ND                \\
          &             & -106.6 $\pm$ 0.3  & 25.3  $\pm$ 2.6   & 84.2 $\pm$ 2.6  & ND                & ND               & ND                \\
 DQ Tau   & 11-02-2019  & -18.9 $\pm$ 2.0   & 63.5  $\pm$ 5.7   & 13.7 $\pm$ 1.0  & ND                & ND               & ND                \\ 
          &             & -78.1 $\pm$ 7.4   & 31.0  $\pm$ 16.1  & 4.1  $\pm$ 1.3  & ND                & ND               & ND                \\ 
 DR Tau   & 01-25-2020  & -19.5 $\pm$ 9.6   & 70.6  $\pm$ 24.7  & 3.3  $\pm$ 0.6  & ND                & ND               & ND                \\   
 DS Tau   & 11-01-2019  & ND                & ND                & ND              & ND                & ND               & ND                \\ 
 FT Tau   & 01-25-2020  & -105.0 $\pm$ 13.0 & 94.2  $\pm$ 20.7  & 1.0  $\pm$ 0.2  & ND                & ND               & ND                \\     
 GG Tau   & 12-09-2018  & -20.0 $\pm$ 8.6   & 13.9  $\pm$ 22.4  & 2.4  $\pm$ 2.3  & -14.8 $\pm$ 7.6   & 5.3 $\pm$ 7.1    & 12.5 $\pm$ 8.8    \\ 
 GH Tau   & 01-25-2020  & -69.0 $\pm$ 12.6  & 47.0  $\pm$ 32.7  & 1.5  $\pm$ 0.6  & ND                & ND               & ND                \\
 GI Tau   & 01-26-2020  & 42.0 $\pm$ 5.9    & 35.2  $\pm$ 13.6  & 2.0  $\pm$ 0.5  & ND                & ND               & ND                \\ 
          &             & -43.0 $\pm$ 7.3   & 62.3  $\pm$ 14.3  & 2.0  $\pm$ 0.4  & ND                & ND               & ND                \\   
 GK Tau   & 11-02-2019  & ND                & ND                & ND              & ND                & ND               & ND                \\          
 GM Aur   & 12-09-2018  & ND                & ND                & ND              & -15.5 $\pm$ 3.8   & 4.2 $\pm$ 2.6    & 4.3 $\pm$ 2.6     \\
 HL Tau   & 11-24-2016  & -                 & -                 & -               & ND                & ND               & ND                \\ 
 HN Tau   & 12-19-2015  & -                 & -                 & -               & ND                & ND               & ND                \\
          & 10-29-2017  & -                 & -                 & -               & ND                & ND               & ND                \\          
 HQ Tau   & 11-01-2019  & ND                & ND                & ND              & ND                & ND               & ND                \\
 IP Tau   & 12-09-2018  & ND                & ND                & ND              & ND                & ND               & ND                \\ 
 IQ Tau   & 11-02-2019  & -30.0 $\pm$ 1.2   & 34.0  $\pm$ 3.0   & 69.5 $\pm$ 1.2  & -11.4 $\pm$ 2.7   & 11.2 $\pm$ 2.5   & 24.5 $\pm$ 1.0    \\            
 MWC 480  & 11-01-2019  & ND                & ND                & ND              & ND                & ND               & ND                \\           
 RW Aur A & 11-13-2017  & 111.6  $\pm$ 7.9  & 73.9  $\pm$ 15.5  & 75.5 $\pm$ 13.7 & ND                & ND               & ND                \\
          &             & -90.9  $\pm$ 8.9  & 211.3 $\pm$ 19.7  & 108.0 $\pm$ 9.8 & -20.8 $\pm$ 8.0   & 15.2 $\pm$ 5.6   & 36.0 $\pm$ 20.2   \\
          &             & -172.1 $\pm$ 9.9  & 73.8  $\pm$ 22.5  & 65.0 $\pm$ 16.3 & -24.0 $\pm$ 13.0  & 179.7 $\pm$ 14.9 & 38.0 $\pm$ 5.4    \\
 RY Tau   & 11-13-2017  & -56.9  $\pm$ 14.1 & 56.9  $\pm$ 15.7  & 25.2 $\pm$ 7.1  & ND                & ND               & ND                \\
 UX Tau   & 01-26-2020  & 23.5   $\pm$ 5.7  & 61.2  $\pm$ 20.3  & 8.8  $\pm$ 1.9  & ND                & ND               & ND                \\ 
 UY Aur   & 12-08-2018  & -20.5  $\pm$ 7.9  & 23.2  $\pm$ 12.3  & 8.4  $\pm$ 4.4  & ND                & ND               & ND                \\
          &             & -30.0  $\pm$ 19.8 & 164.8 $\pm$ 64.2  & 9.5  $\pm$ 2.0  & ND                & ND               & ND                \\     
 UZ Tau E & 12-09-2018  & -72.0  $\pm$ 1.5  & 63.5  $\pm$ 8.1   & 46.0 $\pm$ 6.0  & ND                & ND               & ND                \\          
 V409 Tau & 01-26-2020  & -50.0  $\pm$ 3.8  & 68.2  $\pm$ 10.9  & 2.3  $\pm$ 0.3  & ND                & ND               & ND                \\
          &             & -123.1 $\pm$ 6.5  & 44.7  $\pm$ 21.0  & 0.8  $\pm$ 0.2  & ND                & ND               & ND                \\ 
 V836 Tau & 11-02-2019  & ND                & ND                & ND              & ND                & ND               & ND                \\
 XZ Tau   & 03-14-2017  & -                 & -                 & -               & ND                & ND               & ND                \\	
\hline
\end{tabular}
\begin{quotation}
\textbf{Notes.} ND: profile not-detected, "-": spectrum not-acquired.
\end{quotation}
\end{table*}

\end{appendix}

\end{document}